\documentclass[a4paper,11pt]{article}
\usepackage{amsfonts,amsthm,amsmath,amssymb,graphicx,hyperref,color,subfigure}

\def\bra#1{\langle #1 |}
\def\ket#1{| #1\rangle}
\newcommand{\braket}[2]{\langle #1 | #2 \rangle}

\def\corr#1{\left\langle #1 \right\rangle}
\newcommand{\tr}{\operatorname{tr}}

\newcommand{\Sym}{\operatorname{Sym}}
\newcommand{\Aut}{\operatorname{Aut}}

\def\id{\textrm{id}}

\def\Sym{\textrm{Sym}}
\def\ha{\frac{1}{2}}

\def\a{\alpha}
\def\b{\beta}

\def\e{\epsilon}
\def\ve{\varepsilon}

\def\m{\mu}

\def\n{\nu}

\def\O{\Omega}

\def\s{\sigma}

\def\S{\Sigma}
\def\l{\lambda}

\def\ov{\overline}
\def\C{\mathbb{C}}
\def\CP{\mathbb{CP}}
\def\R{\mathbb{R}}

\def\Z{\mathbb{Z}}
\def\Q{\mathbb{Q}}

\def\bI{\mathbb{I}}

\newcommand{\cM}{\mathcal M}
\newcommand{\cN}{\mathcal N}
\newcommand{\cO}{\mathcal O}

\newcommand{\cT}{\mathcal T}

\newcommand{\cZ}{\mathcal Z}

\newcommand{\be}{\begin{equation}}
\newcommand{\bea}{\begin{eqnarray}}

\newcommand{\ee}{\end{equation}}
\newcommand{\eea}{\end{eqnarray}}

\newcommand{\nn}{\nonumber}

\title{Complex matrix model duality}

\begin{document}

\rightline{DESY 10-140}

\vspace{4truecm}

\centerline{\LARGE \bf Complex matrix model duality}

\vspace{1truecm}

\centerline{{\large \bf T.W. Brown${}^\star$}}

\vspace{.4cm}
\centerline{{\it  DESY Hamburg, Theory Group,}}
\centerline{{\it  Notkestrasse 85, D-22603 Hamburg, Germany.}}

\vspace{1.5truecm}

%%%%%%%%%%%%%%%%%
\thispagestyle{empty}

\centerline{\bf ABSTRACT}

\vspace{.5truecm}

\noindent The same complex matrix model calculates both tachyon
scattering for the $c=1$ non-critical string at the self-dual radius
and certain correlation functions of half-BPS operators in $\cN=4$
super-Yang-Mills.  It is dual to another complex matrix model where
the couplings of the first model are encoded in the Kontsevich-like
variables of the second.  The duality between the theories is mirrored
by the duality of their Feynman diagrams.  Analogously to the
Hermitian Kontsevich-Penner model, the correlation functions of the
second model can be written as sums over discrete points in subspaces
of the moduli space of punctured Riemann surfaces.

\vfill
\noindent{\it ${}^\star$nworbmot@gmail.com}

\newpage

\tableofcontents

\setcounter{footnote}{0}

\section{Introduction}

In recent examples of gauge-gravity duality, Feynman graphs of the
gauge theory are lifted to open string diagrams whose worldsheet holes
are summed over, replacing them with closed string insertions, to give
a closed string theory in a different background.  Open-closed string
dualities like this, such as the 3d Chern-Simons to conifold duality
\cite{9811131,0205297} or the Kontsevich matrix model to 2d
topological gravity duality \cite{Kontsevich:1992ti,0312196}, have
been categorised by Gopakumar as of `$F$ type' \cite{Gopakumar}
because it is the \emph{faces} of Feynman graphs which are replaced by
closed string insertions.  On the other hand dualities such as that
between 4d $\cN=4$ super Yang-Mills and Type IIB closed string theory
on $AdS_5 \times S^5$ \cite{adscft} are of `$V$ type' \cite{Gopakumar}
because it is the \emph{vertices} of the Feynman diagrams, corresponding to
local operators and interaction vertices, which are replaced by closed
string insertions.  All known open-closed dualities are either of $F$
type or $V$ type \cite{Gopakumar}.  The possibility was raised by
Gopakumar \cite{Gopakumar} that for every gauge-gravity duality the
closed string theory has open string duals of both types, related to
each other by graph duality.  Topological gravity in 2d was given as
an example, with the theory of $V$ type being the double-scaled
Hermitian matrix model
\cite{Brezin:1990rb,Douglas:1989ve,Gross:1989vs} and that of $F$ type
being the Kontsevich Hermitian matrix model \cite{Kontsevich:1992ti}.
Using the proof of equivalence in \cite{0408039}, the graph duality
can be shown dynamically by integrating in and out different fields,
so that at different steps vertices are replaced by faces and vice
versa \cite{Gopakumar}.

\vspace{0.4cm}

In this paper we show this open-open duality between a complex matrix
model of $V$ type called the $Z$ model and another complex matrix
model of $F$ type called the $F$ model.  The $Z$ model is known to
calculate certain correlation functions of half-BPS operators in
$\cN=4$ super Yang-Mills as well as tachyon scattering amplitudes for
the $c=1$ string at the self-dual radius.  The dual $F$ model provides
a new Kontsevich-Penner matrix model for these amplitudes and thus
throws new light on how to write them as integrals over the moduli
space of Riemann surfaces.

\vspace{0.4cm}

The $Z$ model is
\begin{equation}
\boxed{  \cZ(\{t\},\{{\ov t}\}) 
 =\int [dZ]^{\C}_{N\times N}\;\; e^{-\tr(ZZ^\dagger) + \sum_{k=1}^\infty t_k \tr(Z^k) +\sum_{k=1}^\infty{\ov t}_k\tr( Z^{\dagger k})}} \label{eq:Zmodel}
\end{equation}
It has two infinite sets of couplings which are often called times in
the literature because of the relation with the $\tau$-function of the
Toda integrable hierarchy.

From the 4d $\cN=4$ super Yang-Mills perspective the $Z$ model is a
generating function for certain correlation functions of holomorphic
and antiholomorphic half-BPS operators built from a single complex
scalar transforming in the adjoint of the gauge group $U(N)$
\cite{0205033,0205089}.  In `extremal' correlation functions, for
which the antiholomorphic operators are all at the same spacetime
position, the spacetime dependence of the correlation function factors
out of the result; the $Z$ model computes the remaining combinatorial
factor, which is an expansion in $1/N$.\footnote{This combinatorial
  factor is unchanged if all the holomorphic operators are also taken to
  the same spacetime position, so really the $Z$ model just generates
  the two-point function of multi-trace half-BPS states.  This is like
  a metric on the multi-trace states, cf. the discussion in
  \cite{1002.2099}.  Note too that the extremal correlation functions
  are known not to renormalise when the coupling is non-trivial.\label{foot:metric}}
Because local operators in $\cN=4$ super Yang-Mills (vertices in the
$Z$ model) map to string (or supergravity in this case) states, the
$Z$ model is of $V$ type.

In the guise of a normal matrix model \cite{0302106} the $Z$
model is also a generating function for the correlation functions of
integer-momentum massless tachyons in the $c=1$ non-critical string
compactified at the self-dual radius. The cosmological constant $\m$
of the $c=1$ string, which controls the genus expansion, is related to
the rank $N$ of the complex matrix by $N= -i \m$.  In contrast to the
double-scaled matrix quantum mechanics (MQM) for the $c=1$
string,\footnote{See \cite{9108019,9304011} for reviews and references
  therein.} the $Z$ model requires no scaling limit and works at
finite $N$.  This is reflected in the fact that the $Z$ model is not a
triangulation of the Riemann surface itself but rather, through its
dual, a triangulation of the moduli space of punctured Riemann
surfaces. This relation between the half-BPS sector of the $AdS_5$
duality and the $c=1$ string has been explored in
\cite{0403110,0612262} based on the similarity of their MQM
descriptions \cite{0111222} and shown to be exact at the self-dual
radius in \cite{0701086}.  This connection is in the spirit of the
minimal $(p,1)$ string embedding in the $AdS_3$ duality
\cite{0507037}.

The map from tachyons $\cT_{p}$ with integer momentum $p$ to matrix
variables is for $k>0$
\begin{equation}
  \cT_k \to \tr(Z^k) \hspace{2cm}  \cT_{-k} \to \tr(Z^{\dagger k})
\end{equation}
The individual tachyon correlation functions are then
\begin{equation}
   \corr{ \cT_{k_1}\cdots \cT_{k_p}\;\; \cT_{-{\ov k}_1}\cdots \cT_{-{\ov k}_q}}_{c=1}  = \corr{ \tr(Z^{k_1}) \cdots \tr(Z^{k_p})\;\; \tr(Z^{\dagger{\ov k}_1}) \cdots \tr(Z^{\dagger{\ov k}_q}) } \label{eq:tachycorr}
\end{equation}
On the righthand side the correlation function is taken using the
complex matrix model with Gaussian action $\tr(ZZ^\dagger)$.  It is computed by Wick-contracting with the
propagator
\begin{equation}
  \corr{Z^e_f \; Z^\dagger {}^g_h} = \delta^e_h \delta^g_f
\end{equation}
These correlation functions can be computed to all orders
in $N$ using symmetric group techniques \cite{0111222,1002.2099}.

The dual $F$ model is exactly the same function of $\{t\},\{{\ov t}\}$
\begin{align}
\boxed{\begin{array}{rl}\cZ(\{t\},\{{\ov t}\}) & = \int [dF]^\C_{n\times n}\;\;e^{-\tr(FF^\dagger) -N\tr \log\left(1- A^{-1}F
    B^{-1}F^\dagger \right)}  \\
& \\
& =  \int [dF]^\C_{n\times n}\;\;e^{-\tr(FF^\dagger) +N \sum_{k=1}^\infty \frac{1}{k} \tr\left[\left( A^{-1}F
    B^{-1}F^\dagger \right)^{k} \right]} \end{array}} \label{eq:Fmodel}
\end{align}
The couplings $\{t\}$ and $\{{\ov t}\}$ are encoded in matrices $A,B$ by a Kontsevich-Miwa transformation
\begin{equation}
   t_k = \sum_{i=1}^n \frac{1}{ka_i^k} = \frac{1}{k} \tr A^{-k} \hspace{2cm} {\ov t}_k = \sum_{j=1}^n \frac{1}{kb_j^k} = \frac{1}{k} \tr B^{-k}
\end{equation}
To expand $\cZ(\{t\},\{{\ov t}\})$ in these variables, compute correlation functions with the even-valency vertices that appear in \eqref{eq:Fmodel}
for $k>1$ using the propagator from the matrix model
\begin{equation}
  \corr{F^i_j \; F^\dagger{}^k_l } = \frac{\delta^i_l \delta^k_j}{(1-Na_i^{-1} b_j^{-1})} \label{eq:Fprop}
\end{equation}
The colour index for each face of the $F$ model Feynman diagrams
comes with either an $a_i$ or a $b_j$, so the couplings $\{t\}$ and
$\{{\ov t}\}$ are associated to faces of the $F$ model.  Thus
the $F$ model is of $F$ type.

Relations between Hermitian matrix models via graph duality have
appeared before in the literature, as have complex matrix models
similar to the $F$ model (see for example
\cite{Morris:1990cq}-\cite{1005.5715}, also in connection with
$\tau$-functions \cite{9303139,math-ph/0210012}).  The $F$ model is of
Kontsevich type because the couplings are encoded and expanded
similarly to the Kontsevich model for topological gravity
\cite{Kontsevich:1992ti}.  It is also of Penner type because the
appearance of a logarithmic term in the action is similar to the
Penner model for the virtual Euler characteristic of the moduli space
of punctured Riemann surfaces $\cM_{g,n}$ \cite{Penner}.  In fact the
$F$ model is a complex matrix model analogue of the Hermitian
Kontsevich-Penner model studied by Chekhov and Makeenko
\cite{9202006}, which is dual to the Hermitian version of the $Z$
model (before the double-scaling limit) in exactly the same way
\cite{9202006,Gopakumar}.

\vspace{0.4cm}

The most direct way to prove the duality between the $Z$ and $F$
models is using character expansions, see Section \ref{sec:halfbps}.
Term-by-term it can be seen that the Feynman diagrams of the different
models are graph-dual.  In Section \ref{sec:Hurwitz} it is shown that
the correlation functions are in fact sums over Hurwitz numbers, which
count holomorphic maps from the worldsheet to $\CP^1$ branched just
three times.  The valencies of the vertices and faces specify the
ramification profiles.

Another proof is given in Section \ref{sec:proof} with the techniques
used in the 2d topological gravity case by Maldacena, Moore, Seiberg,
Shih \cite{0408039} and Gopakumar \cite{Gopakumar}, which involve
integrating fields in and out twice.  In this method the graph duality
between the $Z$ and $F$ model can be seen `dynamically', as explained
in Section \ref{sec:graphdual}.  Every Feynman diagram in the original
$Z$ matrix model corresponds to a diagram in the $F$ matrix model to
which it is dual.  This insight is crucial to read off the correct
terms that are identified in the different models.  In fact with the
propagator \eqref{eq:Fprop} the $F$ model is only sensitive to
`skeleton' graphs of the $Z$ model where propagators running parallel
between the same vertices are bunched together into the same edge.
These skeleton graphs were introduced in \cite{0402063} as part of
Gopakumar's programme to find the closed string duals of free gauge
theories \cite{0308184,0402063,0504229}.

An advantage of the $F$ model is that its correlation functions can be
expressed directly as integrals over the moduli space $\cM_{g,n}$ of
punctured Riemann surfaces, using the example set by the Kontsevich
model \cite{Kontsevich:1992ti}.  In the Schwinger parameterisation of
the propagators, the Schwinger lengths associated to each edge of each
Feynman graph provide coordinates on a cell decomposition of
$\cM_{g,n}$.  The integrals over the top-dimensional cells in
$\cM_{g,n}$ require all vertices of the graphs to be trivalent.  The
vertices of the $F$ model have a minimum valency of four, which means
that the correlation functions can only come from lower-dimensional
cells in the moduli space.  Furthermore, following the analysis of the
Hermitian Kontsevich-Penner model in \cite{9302014}, the integral
localises on discrete points in these subspaces, see Section
\ref{sec:discussion}.

Despite the fact that the $Z$ model needs no double-scaling limit for
its identification with the $c=1$ string, it is still possible to take
one.  From the $\cN=4$ perspective it is the BMN limit \cite{0202021}
and it limits the $F$ model to only 4-valent vertices, see Section
\ref{sec:klimit}.  The meaning of this limit for the $c=1$ string is
unclear.  Rewriting the correlation functions of the $Z/F$ model in
terms of Hurwitz numbers in Section \ref{sec:Hurwitz}, this limit
involves restricting to a special class of Hurwitz numbers called
double Hurwitz numbers, with arbitrary branching profiles at two
points and simple branchings elsewhere.

Another topological matrix model for the $c=1,R=1$ string is the
$W_\infty$ model of \cite{9505127,9208031}, reviewed in
\cite{0310287}, where just the positive momentum tachyon couplings are
rearranged in this way
\begin{equation}
  \cZ(\{t\},\{{\ov t}\}) =  \int [dM]^H_{N\times N} \;\;e^{\tr(-M +  \sum_{k=1}^\infty {\ov t}_k (MA^{-1})^k)}
\end{equation}
The integral is over a Hermitian matrix $M$.  The relation of the $Z$
model to this $W_\infty$ model was explained by Mukherjee and Mukhi in
\cite{0505180}; a direct transformation of the $W_\infty$ model into
the $F$ model is shown in Appendix \ref{WinftoF}.

\section{Proof of duality using integration in-out-in-out}\label{sec:proof}

In this section the duality between the $Z$ and $F$ models is proved
using the techniques of \cite{0408039,Gopakumar} by integrating in and
out different fields.  This makes the graph duality of the models
manifest, as is explained in the next section.

The partition function for the $Z$ model is
\begin{equation}
  \cZ(\{t\},\{{\ov t}\}) = \int [dZ]^\C_{N\times N}\;\; e^{-\tr(ZZ^\dagger) + \sum_{k=1}^\infty t_k\tr( Z^k) +\sum_{k=1}^\infty {\ov t}_k \tr(Z^{\dagger k})}\label{eq:NMM}
\end{equation}
This model is the same as the Model II for the $c=1$ string at the
self-dual radius $R=1$ with $N=\n \equiv -i \m$ in \cite{0302106}.
Although the integration in \cite{0302106} is over a normal matrix
with the condition $[Z,Z^\dagger]=0$ enforced, with this action for
$R=1$ both the complex and normal matrix model are the
same.\footnote{A normal matrix can be decomposed into a unitary matrix
  $U$ and a diagonal matrix of its complex eigenvalues $D$, $Z_N =
  UDU^\dagger$.  This is not true for a complex matrix, for which we
  have $Z = U(D+R)U^\dagger$ where $R$ is strictly upper triangular
  \cite{Mehta}.  It can be checked that in the action \eqref{eq:NMM}
  $R$ completely decouples, and since the measure on $U$ and $D$ is
  the same, the normal matrix model is equivalent to the complex
  matrix model with this action.  See equation
  \eqref{eq:complexdecomp} in Appendix \ref{WinftoF} for an
  alternative way to decompose a complex matrix.}  The tachyon scattering matrix
agrees with older results calculated in the literature
\cite{Klebanov:1991ai}.

Substitute the $t_k$ and ${\ov t}_k$ for two diagonal $n \times n$
matrices $A$ and $B$, with eigenvalues $a_i$ and $b_j$ respectively,
using the Kontsevich-Miwa transformation
\begin{equation}
   t_k = \frac{1}{k} \tr A^{-k} = \sum_{i=1}^n \frac{1}{ka_i^k} \hspace{2cm} {\ov t}_k= \frac{1}{k} \tr B^{-k} = \sum_{j=1}^n \frac{1}{kb_j^k}  \label{eq:KMid}
\end{equation}
For the $t_k$ to be independent whenever the $\tr(Z^k)$ are, we
need $ n \geq N$ and similarly for the ${\ov t}_k$.

The exponentiated $\tr(Z^k)$ operators can be written as inverse
determinants provided the $a_i$ are sufficiently large (to avoid convergence issues)
\begin{align}
&\exp \left[ \sum_{k=1}^\infty t_k\tr (Z^k)  \right]  =\exp \left[ \sum_{k=1}^\infty\sum_{i=1}^n \frac{1}{ka_i^k}\tr (Z^k)  \right] \nn\\
&\hspace{4cm} = \exp \left[-\sum_{i=1}^n \tr\log\left(1- \frac{Z}{a_i}\right)  \right] = \prod_{i=1}^n \left[\det\left(1-\frac{Z}{a_i}\right)\right]^{-1} \label{eq:expandAB}
\end{align}
[In the 2d topological gravity case the determinants in the double-scaled
  Hermitian matrix model correspond to exponentiated macroscopic loop
  operators for FZZT branes \cite{FZZT}
\begin{equation}
  \tr \log (a_i - M) = \int \frac{dl}{l}\tr e^{-l(a_i -M)} 
\end{equation}
Each of the $n$ FZZT branes has boundary cosmological
constant $a_i$.  There is no clear such interpretation of the
determinants as wavefunctions of FZZT branes here, and in fact the
more natural extension of \cite{0408039} would be to investigate
macroscopic loops in the matrix quantum mechanics, cf. \cite{0503112}.
The fact that we have \emph{inverse} determinants in the $c=1$ case
(also present in the study of the normal matrix model in
\cite{0505180}) also differs from \cite{0408039} and alters the
statistics for the fields that we integrate in later, which in the
$c<1$ case are fermionic strings stretching between the ZZ \cite{0101152} and FZZT
branes.\footnote{Note that if we had chosen to include a minus sign in
  the identification \eqref{eq:KMid} we would have had normal
  determinants here and fermions integrated in later.  The choice of
  sign is left to a physical interpretation in the future.} From the
4d $\cN=4$ SYM perspective, these determinants (more clearly expanded
in equation \eqref{eq:expexpansion}) are interpreted as giant graviton
branes in the bulk \cite{0107119,0111222}.]

Using \eqref{eq:expandAB} the $Z$ model partition function is now
\begin{align}
  \cZ(\{t\},\{{\ov t}\})
& =   \int [dZ]^\C_{N\times N}\;\; e^{-\tr(ZZ^\dagger) + \sum_{k=1}^\infty t_k \tr(Z^k) +\sum_{k=1}^\infty{\ov t}_k \tr(Z^{\dagger k})} \nn \\
 & = \det(A)^{N}\det(B)^{N}  \corr{ \prod_{i=1}^n \frac{1}{\det(a_i-Z)} \;\;\prod_{j=1}^n \frac{1}{\det(b_j-Z^\dagger)} }  \label{eq:detstoNMM}
\end{align}
The correlation function is taken with the Gaussian action
$\tr(ZZ^\dagger)$, as will always be the case for the $Z$ model.

Writing the products of determinants using single determinants of larger $nN\times nN$ matrices, we can write them as integrals over two sets of complex bosonic fields\footnote{This type of identity for the determinants only works with the determinant of a Hermitian matrix.  We extend it to our case by noting that the $Z$ model only depends on the eigenvalues of $Z$ and since they are complex we can extend the identity by analytic continuation.}
\begin{align}
& \corr{\frac{1}{\det( A \otimes \bI_N - \bI_n \otimes Z)}\frac{1}{\det( B \otimes \bI_N - \bI_n \otimes Z^\dagger)}} \nn \\
& =  \int [dZ]^\C_{N \times N}[dC]^\C_{N \times n}[dD]^\C_{N \times n}\;\; e^{-\tr\left[ZZ^\dagger + C^\dagger( A \otimes \bI_N - \bI_n \otimes Z)C + D^\dagger( B \otimes \bI_N - \bI_n \otimes Z^\dagger)D \right]} \label{eq:ZCD}
\end{align}
$C_{ei}$ and $D_{ej}$ are bifundamental fields with $e=1, \dots N$ and
$i,j=1, \dots n$.  Again because we have inverse determinants this
contrasts to the minimal string case, in which one must integrate in
fermions rather than bosons.

Next integrate out the $Z$ field, after rewriting \eqref{eq:ZCD} appropriately
\begin{align}
& \int [dZ][dC][dD]\;\; e^{-\tr\left[(Z-DD^\dagger)(Z^\dagger -CC^\dagger ) -CC^\dagger DD^\dagger + C^\dagger AC + D^\dagger BD\right] } \nn \\
& =  \int[dC][dD]\;\; e^{ -\tr\left(C^\dagger A C +  D^\dagger B D  - CC^\dagger DD^\dagger\right)} \label{eq:CD}
\end{align}
This is the $C,D$ matrix model.  The quartic vertex is $CC^\dagger DD^\dagger
= C_{ei} C^\dagger_{fi} D_{fj}D^\dagger_{ej}$.  It has propagators
\begin{equation}
  \corr{C_{ei_1}C^\dagger_{fi_2}} = \frac{\delta_{ef}\delta_{i_1i_2}}{a_{i_1}} \hspace{1.5cm}  \corr{D_{ej_1}D_{fj_2}^\dagger} = \frac{\delta_{ef}\delta_{j_1j_2}}{b_{j_1}} \label{eq:CDprop}
\end{equation}
The $C,D$ model can also be expanded as a function of the couplings
$\{t\},\{{\ov t}\}$, see Appendix \ref{sec:CDexamples}.

Next we integrate back in an $n\times n$ complex matrix $F^i_{j}$ being careful with the indices
\begin{align}
&  \int [dC][dD]\;\; e^{- C^\dagger_{ei_1} A_{i_1i_2}C_{ei_2} -  D^\dagger_{ej_1} B_{j_1j_2}D_{ej_2} +C_{ei}D^\dagger_{ej}D_{fj} C^\dagger_{fi}}\nn\\
& =   \int [dF]^\C_{n\times n} [dC][dD]\;\; e^{-(F^i_{j} -C_{ei} D^\dagger_{ej})(F^\dagger{}^j_{i} - D_{fj}C^\dagger_{fi} ) + C_{ei}D^\dagger_{ej}D_{fj} C^\dagger_{fi}- C^\dagger_{ei_1} A_{i_1i_2}C_{ei_2} -  D^\dagger_{ej_1} B_{j_1j_2}D_{ej_2} }\nn\\
&= \int [dF]^\C_{n\times n}[dC][dD]\;\;e^{-\tr\left(FF^\dagger - D^\dagger F^\dagger C - C^\dagger F D + C^\dagger AC +  D^\dagger BD \right)}  \label{eq:FCD}
\end{align}
To integrate out $C$ and $D$ write them together as a single $N\times 2n$ field so that the cubic terms become
\begin{equation}
  \left(
  \begin{array}{cc}
    C^\dagger &
    D^\dagger
  \end{array}
  \right)
  \left(
  \begin{array}{cc}
    A & -F \\
    -F^\dagger & B
  \end{array}
  \right)
  \left(
  \begin{array}{c}
    C \\
    D
  \end{array}
  \right)
\end{equation}
The result is an inverse determinant of an $[N\times 2n]\times [N\times 2n]$ matrix
\begin{equation*}
 \int [dF]\left\{\det\left[  \left(
  \begin{array}{cc}
    A & -F \\
    -F^\dagger & B
  \end{array}
  \right)\otimes \bI_N \right]\right\}^{-1} e^{-\tr(FF^\dagger) }  = \int [dF] \left\{\det \left(
  \begin{array}{cc}
    A & -F \\
    -F^\dagger & B
  \end{array}
  \right)\right\}^{-N} e^{-\tr(FF^\dagger) }
\end{equation*}
Now write the determinant in terms of a product of matrices
\begin{align}
 \det \left(
  \begin{array}{cc}
    A & -F \\
    -F^\dagger & B
  \end{array}
  \right)  
& =  \det \left( \left(
  \begin{array}{cc}
    A & 0 \\
    0 & B
  \end{array}
  \right)\left(
  \begin{array}{cc}
    1 & -A^{-1}F \\
    -B^{-1}F^\dagger & 1
  \end{array}
  \right) \right) \nn \\
& =  \det(A) \det(B) \det\left(
  \begin{array}{cc}
    1 & -A^{-1}F \\
    -B^{-1}F^\dagger & 1
  \end{array}
  \right) 
\end{align}
The constant terms $\det(A)^{-N} \det(B)^{-N}$ cancel those in \eqref{eq:detstoNMM}, so we get
\begin{align}
 \cZ(\{t\},\{{\ov t}\}) & =  \int [dF]\;\;e^{-\tr(FF^\dagger) - N\tr \log\left(
  \begin{array}{cc}
    1 & -A^{-1}F \\
    -B^{-1}F^\dagger & 1
  \end{array}
  \right) } 
\end{align}
Expanding the logarithm for large $A,B$, just as we did in \eqref{eq:expandAB}, we find
\begin{align}
 \cZ(\{t\},\{{\ov t}\}) & =  \int [dF]^\C_{n\times n}\;\;e^{-\tr(FF^\dagger) + N \sum_{k=1}^\infty \frac{1}{k} \tr\left[\left(
  \begin{array}{cc}
    0 & A^{-1}F \\
    B^{-1}F^\dagger & 0
  \end{array}
  \right)^k \right]}  \nn \\
& =  \int [dF]^\C_{n\times n}\;\;e^{-\tr(FF^\dagger) +N \sum_{k=1}^\infty \frac{1}{k} \tr\left[\left( A^{-1}F
    B^{-1}F^\dagger \right)^{k} \right]}  \label{eq:FFdagger}
\end{align}
This is the $F$ model introduced in equation \eqref{eq:Fmodel}. To extract the propagator study the quadratic term
\begin{equation}
   F^i_j \left( \delta^l_i \delta^j_k - N(A^{-1})^l_i (B^{-1})^j_k \right) F^\dagger{}^k_l
\end{equation}
Using $(A^{-1})^l_i = a_i^{-1}\delta^l_i$ the propagator is
\begin{equation}
  \corr{F^i_j \; F^\dagger{}^k_l } = \frac{\delta^i_l \delta^k_j}{(1-Na_i^{-1} b_j^{-1})} \label{eq:FFprop}
\end{equation}

Alternatively we could have taken the plain quadratic term
$\tr(FF^\dagger)$ with plain propagator
\begin{equation}
  \corr{F^i_j \; F^\dagger{}^k_l }_{\textrm{plain}} = \delta^i_l
  \delta^k_j \label{eq:FFplain}
\end{equation}
and treated the $k=1$ term
$N\tr(A^{-1}FB^{-1}F^\dagger)$ from \eqref{eq:FFdagger} as an additional
 interaction vertex.  The propagator
\eqref{eq:FFprop}  is then a sum over an arbitrary number of 
intervening such 2-valent vertices
\begin{equation}
  \sum_{p=0}^\infty  \frac{N^p}{p!}\corr{F^i_j \;\left[\tr(A^{-1}FB^{-1}F^\dagger)\right]^p \;F^{\dagger}{}^k_l    }_{\textrm{plain}}  = \sum_{p=0}^\infty (Na_{i}^{-1}b_j^{-1})^p\; \delta^i_l \delta^k_j = \frac{\delta^i_l \delta^k_j}{(1-Na_i^{-1} b_j^{-1})} \label{eq:bigsum}
\end{equation}
This is an important issue for the interpretation of the graph duality
in Section \ref{sec:graphdual}, since these 2-valent vertices are
exactly those which are dual to the faces bounded by parallel
propagators between the same vertices, see Figure
\ref{fig:oomm-parallel}.  Bunching such parallel propagators into a
single edge, reducing the Feynman diagram to a skeleton graph, removes
such 2-valent vertices from the dual graph.  This works because the
propagator \eqref{eq:FFprop} sums over all possible $Z$ diagrams with
the same skeleton graph.  These interpretational issues are also
crucial to understand the dual of the planar two-point function of the
$Z$ model and the character expansion to which we turn in Section
\ref{sec:halfbps}.

It will often be useful for calculations to rescale the $F$ model $F \to \sqrt{A}F\sqrt{B}$ to get
\begin{equation}
\cZ(\{t\},\{{\ov t}\}) = (\det A)^n (\det B)^n
  \int [dF]^\C_{n\times n}\;\;e^{-\tr(AFBF^\dagger) +N \sum_{k=1}^\infty \frac{1}{k} \tr\left( F
    F^\dagger \right)^{k} }  \label{eq:Fscaled}
\end{equation}
This model then has propagator
\begin{equation}
  \corr{F^i_j \; F^\dagger{}^k_l }_{\textrm{scaled}} = \frac{\delta^i_l \delta^k_j}{a_i b_j-N} \label{eq:Fscaledprop}
\end{equation}
This form is useful for transferring the Hermitian Kontsevich-Penner
analysis of \cite{9302014} to the $F$ model in Section
\ref{sec:discussion}.  Example $F$ model correlation functions are
computed in Appendix \ref{sec:examples}.

\section{Dynamical graph duality}\label{sec:graphdual}

[All figures are printed at the end of the paper to avoid cluttering
  the text.]

In this section the duality between the $Z$ model and the $F$ model is
shown to work at the level of individual Feynman diagrams.  An
explicit proof with formulae is provided in Section \ref{sec:halfbps}.

Each graph of the $Z$ model corresponds to a graph of the $F$ model
which is dual to the original $Z$ model graph.  The graph duality is
shown `dynamically' in the sense that it is split up into stages where
we first replace the vertices of the $Z$ model by faces (integrating
in $C,D$), then contract and expand propagators in different channels
(integrating out $Z$ and in $F$) and then finally replacing the faces
of the $Z$ model by vertices of the $F$ model (integrating out $C,D$).
This analysis completely mirrors the 2d topological gravity analysis
by Gopakumar \cite{Gopakumar}.

We start with a correlation function such as \eqref{eq:tachycorr} for
the $Z$ matrix model with several holomorphic vertices $\tr(Z^k)$ and
several antiholomorphic vertices $\tr(Z^{\dagger k})$.  Each possible
Wick contraction with the propagator will give a different Feynman
graph.  Because the only non-trivial propagator is $Z \to Z^\dagger$
and no vertices mix $Z$'s with $Z^\dagger$'s, there can be no
propagators connecting a vertex back to itself.  The Feynman diagram
has a minimum genus surface on which it can be drawn with no lines
crossing.  Propagators that run parallel to each other between the
same pair of vertices will be refered to as `homotopic' and can be
bunched together into a single edge.  If we do this for all possible
propagators then we get what Gopakumar named a skeleton graph
\cite{0402063}.  For each correlation function there may be several
topologically distinct skeleton graphs, each corresponding to a number
of possible Wick contractions of the original correlation function.
Later in Section \ref{sec:halfbps} we will see that these distinct skeleton
graphs correspond to the cut-and-join operators of \cite{1002.2099}.

It should also be noted that some graphs, where there is more than one
single trace for both holomorphic and antiholomorphic operators, are
disconnected; cf. the examples in Appendix \ref{sec:m04non}.

When we integrate in the $C,D$ matrices, a vertex like $\tr(Z^m)$ is
replaced by a face of $C$'s and an antiholomorphic vertex
$\tr(Z^{\dagger m})$ is replaced by a face of $D$'s, see Figure
\ref{fig:oomm-inCD}.  In the figures the double lines of the $Z$ are
drawn with solid lines $e=1,\dots N$, while the bifundamental $C$'s
involve both a solid line and a dashed line for $i =1,\dots n$ and the
$D$'s a solid line and a dotted line for $j=1,\dots n$.  We get cubic
couplings $C^\dagger Z C$ and $D^\dagger Z^\dagger D$ from the action
in \eqref{eq:ZCD}.
\begin{figure}[p]
\begin{center}
%left, bottom, right, top
\resizebox{!}{6cm}{\includegraphics[trim=0 0 0 0 ]{oomm-inCD.eps}}
\caption[]{$C,D$ integrated in.}\label{fig:oomm-inCD}
\end{center}
\end{figure}

We then integrate out $Z,Z^\dagger$.  The $Z\to Z^\dagger$ propagators
shrink to give graphs with the quartic vertex $CC^\dagger DD^\dagger$,
cf. Figure \ref{fig:oomm-outZ}.  This is the $C,D$ matrix model from
\eqref{eq:CD}.  It can be expanded in its own right, see the examples in Appendix \ref{sec:CDexamples}.
\begin{figure}[h]
\begin{center}
%left, bottom, right, top
\resizebox{!}{1.3cm}{\includegraphics[trim=0 0 0 0 ]{oomm-outZ.eps}}
\caption[]{$Z$ integrated out.}\label{fig:oomm-outZ}
\end{center}
\end{figure}

Integrating in the complex matrix $F$, the quartic vertex $CC^\dagger
DD^\dagger$ expands in a different channel into an $F\to F^\dagger$
propagator with the cubic couplings $C^\dagger F D$ and $D^\dagger
F^\dagger C$ of \eqref{eq:FCD} at each end, cf. Figure
\ref{fig:oomm-inF}.  In this way each propagator of the $Z$ model corresponds to a transverse propagator of the $F$ model via the quartic $CC^\dagger DD^\dagger$ vertex.
\begin{figure}[h]
\begin{center}
%left, bottom, right, top
\resizebox{!}{2.7cm}{\includegraphics[trim=0 0 0 0 ]{oomm-inF.eps}}
\caption[]{$F$ integrated in.}\label{fig:oomm-inF}
\end{center}
\end{figure}

Finally we integrate out $C$ and $D$ to get the dual graph in terms of
$F,F^\dagger$ for the $F$ model \eqref{eq:FFdagger}.  Each face
involving a solid line, corresponding to the original faces of the $Z$
model Feynman diagram, is replaced by an even-valency $F,F^\dagger$
vertex $(FF^\dagger)^p$, cf. Figure \ref{fig:oomm-outCD}.  Faces to
which only two $F \to F^\dagger$ propagators connect become just a
single $F\to F^\dagger$ propagator.  These faces correspond to the
faces between parallel homotopic propagators of the $Z$ model,
cf. Figure \ref{fig:oomm-parallel}.  Thus a bunch of parallel
propagators from the $Z$ model become just a single $F$ propagator in
the dual graph; in other words the $F$ model is sensitive only to the
topology of the skeleton graph.
\begin{figure}[h]
\begin{center}
%left, bottom, right, top
\resizebox{!}{5cm}{\includegraphics[trim=0 0 0 0 ]{oomm-outCD.eps}}
\caption[]{$C,D$ integrated out; note that you can have any even-valency $(FF^\dagger)^p$ of $F,F^\dagger$ vertices.}\label{fig:oomm-outCD}
\end{center}
\end{figure}

We have thus seen that the correspondence between the $Z$ model and
the $F$ model corresponds to graph duality.  A vertex of the $Z$ model
becomes a face of the $F$ model; homotopically bunched propagators
(edges) of the $Z$ model become a single propagator (edge) of the $F$
model which is perpendicular to the original $Z \to Z^\dagger$
propagators; and finally the faces of the $Z$ model become
even-valency vertices of the $F$ model.

An important constraining feature of the $F$ model \eqref{eq:FFdagger}
is that the faces of the graphs are always associated with either
$A$'s or $B$'s but never both.  From the propagator in
\eqref{eq:FFprop} the $a_i$ is associated to one (dashed) index line
while the $b_j$ is associated to the other (dotted) index line,
cf. the left part of Figure \ref{fig:oomm-prop}.  The vertices also
preserve the $a_i$ and $b_j$ associations, cf. the right part of
Figure \ref{fig:oomm-prop}.  This is because the faces correspond to
vertices of the $Z$ model where the $A$'s map to $\tr(Z^k)$ vertices
and the $B$'s to $\tr(Z^{\dagger k})$ vertices.  In fact, because a
$Z$ model vertex can never have a self-contraction, each edge of the
$F$ model has an $A$ face on one side and a $B$ face on the other,
reflected in the fact that the double-line propagator for $F$ has one
dashed and one dotted line.
\begin{figure}[h]
\begin{center}
%left, bottom, right, top
\resizebox{!}{3cm}{\includegraphics[trim=0 0 0 0 ]{oomm-prop.eps}}
\caption[]{Associations of $A$ with dashed faces and $B$ with dotted faces.  On the left is the propagator and
  associated $a_i,b_j$.  On the right is an example six-valent vertex
  $\tr\left[(A^{-1}F B^{-1}F^\dagger )^3 \right]$, also with
  associated $a_i,b_j$.}\label{fig:oomm-prop}
\end{center}
\end{figure}

A full example of this dynamical graph duality is given for $\corr{\tr(Z^2) \tr(Z) \tr(Z^{\dagger 3})}$
in Figure \ref{fig:oomm-graphd}.
\begin{figure}[t]
\begin{center}
%left, bottom, right, top
\resizebox{!}{11cm}{\includegraphics[trim=0 0 0 0 ]{oomm-graphd.eps}}
\caption[]{$\corr{\tr(Z^2) \tr(Z) \tr(Z^{\dagger 3})}$ graph duality, step by step.  Graph $(1)$ is the original $Z$ model double-line diagram with three vertices and three propagators.  In $(2)$ $C$ and $D$ are integrated in, replacing the vertices of $Z$ with faces of $C$ and $D$.  In $(3)$ the $Z$ propagators are shrunk to give the quartic vertices $CC^\dagger DD^\dagger$ of the $C,D$ model.  In $(4)$ the quartic vertices are expanded in a different channel as propagators of $F$.  $(5)$ is the same graph as $(4)$, just redrawn on the sphere so that the outer solid line of $(4)$ becomes the inside central solid line of $(5)$.  In $(6)$ $C$ and $D$ are integrated out; the faces of solid lines in $(5)$ have been  replaced by $F/F^\dagger$ vertices.  Now faces of $(6)$ bounded by dashed lines are associated with $A$'s (corresponding to holomorphic vertices of $(1)$) while faces bounded by dotted lines are associated with $B$'s (corresponding to antiholomorphic vertices of $(1)$).}\label{fig:oomm-graphd}
\end{center}
\end{figure}

To conclude this section we summarise the procedure for seeing the
duality from graphs:
\begin{itemize}
\item Starting from the $Z$ model, expand the partition function in
  correlation functions of holomorphic and antiholomorphic operators.
  Each correlation function corresponds to a sum of different
  topologically-distinct skeleton graphs.  Each skeleton graph
  corresponds to a sum of topologically-identical $F$ model Feynman
  diagrams, to which the skeleton graph is dual.
\item Starting from the $F$ model, expand the partition function in
  correlation functions of the interaction vertices.  Each $F$
  correlation function splits into classes of topologically-identical
  Feynman diagrams. Each such class of Feynman diagrams is dual to
  a family of topologically-identical $Z$ skeleton graphs.
\end{itemize}

We will now give an alternative proof of this graph duality using
symmetric group techniques.

\section{Character expansions}\label{sec:halfbps}

In this section we use symmetric group and representation theory
techniques to expand both the $Z$ and $F$ models and show that they
are equal.  The graph duality is revealed in the character expansions
of the models.  In fact a more general correspondence is proven
\begin{align}
   \cZ(\{t\},\{{\ov t}\},\{s \}) 
 = &\int [dZ]^{\C}_{N\times N}\;\; e^{-\tr(S^{-\frac{1}{2}}ZS^{-\frac{1}{2}}Z^\dagger) + \sum_{k=1}^\infty t_k \tr(Z^k) +\sum_{k=1}^\infty{\ov t}_k\tr( Z^{\dagger k})} \label{eq:expansion1} \\
 = & \sum_{l(R) \leq N} \frac{|R|!\;\chi_R(S) \chi_R(A^{-1}) \chi_R(B^{-1})}{d_R} \label{eq:expansion2} \\
 =  &\int [dF]^{\C}_{n\times n} \;\;e^{-\tr FF^\dagger +\sum_{k=1}^\infty s_k \tr(A^{-1}FB^{-1}F^\dagger)^k} \label{eq:expansion3} \\
 \propto & \int[dF]^{\C}_{n\times n} \;\;e^{-\tr(A FBF^\dagger) +\sum_{k=1}^\infty s_k \tr(FF^\dagger)^k} \label{eq:expansion4}
\end{align}
In the first line the quadratic term of the $Z$ model has been
modified with an $N\times N$ diagonal matrix $S$ whose eigenvalues are
$S_1 \cdots S_N$.  The propagator is now
\begin{equation}
  \corr{Z^e_f \; Z^\dagger {}^g_h} = S^{\frac{1}{2}}_e S^{\frac{1}{2}}_f \;\delta^e_h \delta^g_f
\end{equation}
Each face of the $Z$ model is bounded by $2k$ edges (the face has
\emph{valency} $2k$).  Since each edge picks up a factor of
$S^{\frac{1}{2}}$ from the propagator, each $2k$-valent face of the $Z$ model
comes with a factor $\tr(S^{k})$.  Each vertex of the $Z$ model has a
coupling $t_k$ or ${\ov t}_k$ depending on whether it is holomorphic
or antiholomorphic.  To recover the vanilla $Z$ model of 
\eqref{eq:Zmodel} set $S$ to be the identity matrix $S = \bI_N$.

In the second line the $Z$ model has been expanded in characters $\chi_R$ of
the group $U(N)$ whose representations $R$ are labelled by Young
diagrams with at most $N$ rows, $l(R) \leq N$.  $|R|$ is the number of
boxes in the Young diagram and $d_R$ is the dimension of the symmetric
group representation for the same Young diagram.  If we set $S =
\bI_N$ here, the character of $S$ becomes the dimension of the $U(N)$
representation $R$, $\chi_R(\bI_N) = \dim_N R$.  Note the symmetry of
the character expansion in the variables $\{t\}$, $\{{\ov t}\}$ and
$\{s \}$, as encoded in the matrices $A^{-1}$, $B^{-1}$ and $S$.

In the third line this character expansion is equal to the $F$ model,
which has been modified so that now each $2k$-valent vertex is
weighted with a coupling $s_k$.  This coupling is related to the
$N\times N$ matrix $S$ by the Kontsevich-Miwa transformation $s_k \equiv
\frac{1}{k}\tr(S^k)$.  If we set $S = \bI_N$ here, $s_k$ becomes $s_k =
\frac{1}{k} \tr(\bI_N^k) = \frac{N}{k}$ and we recover the vanilla $F$
model of \eqref{eq:Fmodel}.

In the fourth line the $F$ model has been rescaled $F \to
\sqrt{A}F\sqrt{B}$ as in equation \eqref{eq:Fscaled} to make the
duality with the $Z$ model in \eqref{eq:expansion1} more transparent.

In the $Z$ model each $2k$-valent face picks up a coupling $s_k$ from
the propagator while the $k$-valent vertices pick up couplings $t_k$
or ${\ov t_k}$.  On the other hand, for the dual $F$ model, exactly
the reverse happens: it is the $2k$-valent vertices which pick up the
couplings $s_k$ while the $k$-valent faces pick up couplings $t_k$ or
${\ov t_k}$ from the propagator.  Thus the addition of the matrix $S$
and its associated couplings $s_k$ allow us to follow the graph
duality more clearly.

Similar Hermitian models where faces of their graph expansions are
weighted separately from the vertices have been studied in
\cite{9212108}-\cite{9601069} and indeed the $F$ model and its
character expansion were mentioned in the review \cite{9601153} as a
complex generalisation.

\subsection{Character expansion of $Z$ model}

In this section the symmetric group techniques of
\cite{0111222,1002.2099} are used to expand the $Z$ model, which
generates the correlation functions computed in those papers.

The first step is to write each product of holomorphic traces (often
referred to as multi-trace operators) using a conjugacy class element
of $S_k$ where $k$ is the sum of the powers of $Z$.  We write the
conjugacy class as a partition $[\m_1,\m_2, \cdots \m_p]$ of $k$ so
that $k = \sum_i \m_i \equiv |\m|$.\footnote{Assume the parts of the
  partition are ordered $\m_{i} \geq \m_{i+1}$ so that the parts map
  to the row lengths of a Young diagram with $k$ boxes.}  Each $\m_i$
corresponds to a $\m_i$-cycle in $S_k$.  In this way
\begin{equation}
  \tr(Z^{\m_1})   \tr(Z^{\m_2}) \cdots   \tr(Z^{\m_p}) = Z^{e_1}_{e_{\a(1)}}Z^{e_2}_{e_{\a(2)}} \cdots Z^{e_k}_{e_{\a(k)}} \equiv \tr(\a\; Z^{\otimes k}) \label{eq:multi-trace}
\end{equation}
where $\a$ is in the conjugacy class $[\m_1,\m_2, \cdots \m_p]$.  For
a partition $\m$ of $k$ (written $\m \vdash k$) we write a
representative of the corresponding conjugacy class $[\m]$ as $\a_\m
\in [\m] \subset S_k$.  In fact for these operators it does not matter
which representative we pick because all elements in the conjugacy
class $[\m]$, given by $\rho^{-1}\a_\m\rho$ for $\rho \in S_k$,
correspond to the same multi-trace operator \eqref{eq:multi-trace}.
The size of the conjugacy class is 
\begin{equation}
  |[\m]| =
\frac{k!}{|\Sym([\m])|} = \frac{k!}{\prod_{p=1}^k
  p^{i_p(\m)}i_p(\m)!}\label{eq:factors}
\end{equation}
$|\Sym([\m])|$ is the size of the symmetry group $\Sym([\m])$ of the
conjugacy class.  $\Sym(\a_\m)$ is the subgroup of $S_k$ that leaves
$\a_\m$ invariant under conjugation: $\Sym(\a_\m) = \{ \rho \in S_k\;|
\;\rho\a_\m\rho^{-1} = \a_\m\}$.  $i_p(\m)$ is the number of parts
of $\m$ of length $p$.  The factors in the denominator of
\eqref{eq:factors} come from the cyclic symmetry $\cong \Z_p$ of the $i_p(\m)$ cycles of length $p$ and the $i_p(\m)!$ ways of
exchanging them.

Summing over all permutations of Wick contractions of the fields, the
correlation functions of the vanilla $Z$ model
\eqref{eq:Zmodel} are an expansion in $N$ where the power of $N$ is the number of faces of the $Z$ model graph \cite{0111222}
\begin{equation}
  \corr{ \tr(\a_\n \;Z^{\dagger \otimes k}) \;\; \tr(\a_\m \;Z^{\otimes k})}  = 
\sum_{\tau \in S_k} N^{C( \a_\m \tau\a_\n \tau^{-1})} = \sum_{\s,\tau\in S_k} N^{C(\s)} \delta(\s  \a_\m \tau\a_\n \tau^{-1})  \label{eq:CJR}
\end{equation}
$C(\s)$ is the number of cycles in the permutation $\s \in S_k$
(i.e. the number of parts of the corresponding partition).  The second
equality is just a rewriting of the first with the symmetric group
delta function $\delta(\s)$.  $\delta(\s)$ is zero on all elements of
$S_k$ except the group identity $\id$, on which it is 1.

Since the operators only depend on the conjugacy of $\a_\m$ it is
often useful to write them with sums over the entire conjugacy class $\tr(\a_\m \;Z^{\otimes k}) = \frac{1}{|[\m]|}\tr(
\S_\m\;Z^{\otimes k})$ where
\begin{equation}
  \S_\m \equiv \sum_{\a \in [\m]\subset S_k} \a
\end{equation}

The sum over $\s\in S_k$ in \eqref{eq:CJR} can also be sub-divided
into conjugacy classes
\begin{equation}
  \sum_{\s\in S_k} N^{C(\s)}\;\s  =  \sum_{\l\vdash k} N^{C([\l])}\;\S_\l
\end{equation}
This element of the group algebra is often written in the literature
as
\begin{equation}
  \O_k \equiv N^{-k} \sum_{\l\vdash k} N^{C([\l])}\;\S_\l = 
 \sum_{\l \vdash k} \frac{1}{N^{T([\l])}}\; \S_\l \label{eq:Onexp}
\end{equation}
We have used the identity $T([\l]) = k-C([\l])$, where $T([\l])$ is
the minimum number of transpositions needed to build an
element in the conjugacy class $[\l]$.

The correlation function \eqref{eq:CJR} can now be written
\begin{equation}
  \corr{ \tr(\a_\n \;Z^{\dagger \otimes k}) \;\; \tr(\a_\m \;Z^{\otimes k})}  = \frac{k! N^k}{|[\m]|\,|[\n]|} \delta(\S_\n \O_k \S_\m) =\frac{k!}{|[\m]|\,|[\n]|} \sum_{\l \vdash k }N^{C([\l])}\;  \delta(\S_\n \S_\l \S_\m) \label{eq:corrdelta}
\end{equation}
Each summand in \eqref{eq:corrdelta} now corresponds to topologically
different graphs.  The partitions $\m,\n$ label the different
holomorphic and antiholomorphic vertices respectively.  The partition
$\l$, on the other hand, labels the faces.  Each part $\l_i$ of the
partition labels a face so that $C([\l])$ is the total number of faces
and hence gives the power of $N$.  Putting a non-trivial matrix $S$ into the $Z$ model as in \eqref{eq:expansion1} the factor $N^{C([\l])}$ is refined
\begin{equation}
  N^{C([\l])} \to \tr( \a_\l\; S^{\otimes k}) = \prod_{p=1}^k \left[\tr(S^p)\right]^{i_p(\l)}
\end{equation}
$i_p(\l)$ is the number of parts in $\l$ of length $p$.  Each part of
length $p$ corresponds to a $2p$-valent face of the $Z$ graph bounded
by $2p$ edges.  The parts of length $1$ correspond to faces with only
two edges; these are the faces between propagators running parallel
between two vertices, see the lefthand diagram in Figure
\ref{fig:oomm-parallel}.  Bunching these parallel propagators into
single edges so that we get a skeleton graph corresponds to ignoring
the parts of $\l$ of length 1.  This proves a conjecture in
\cite{1002.2099} that the different $\S_\l$ correspond to the
different skeleton graphs if you ignore the parts of length 1. The
genus of the Feynman diagram can be read off using the Euler characteristic with $V = C([\m]) + C([\n])$ vertices, $E = k$ edges and $F = C([\l])$ faces
\begin{equation}
  \chi = 2-2g = V-E+F = C([\m]) + C([\n]) + C([\l]) - k \label{eq:Eulerchar}
\end{equation}

In \cite{1002.2099} the $\S_\l$ were called cut-and-join operators
because they have an action on a multi-trace operator $\tr(\S_\m
\;Z^{\otimes k})$ that can split single traces into many traces or
join many into one. This action is simply left-multiplication
\begin{equation}
  \S_\l: \tr(\S_\m\; Z^{\otimes k}) \mapsto \tr(\S_\l\S_\m\; Z^{\otimes k})
\end{equation}
For example the conjugacy class of transpositions $\l = [2,1^{k-2}]$
gives the cut-and-join operator $\S_{[2]}$.  Acting on the single
trace operator given by $\m = [k]$ splits it into all possible
double-traces $[p,k-p]$
\begin{equation}
  \S_{[2]} \tr(Z^k) = \frac{1}{|[k]|}\tr( \S_{[2]} \S_{[k]}\;Z^{\otimes k}) =  \frac{k}{2}\sum_{p=1}^{k-1} \tr(Z^{p})\tr(Z^{k-p})
\end{equation}
Acting on a double trace with $\S_{[2]}$ can also join it into a
single trace.  Similar results follow when acting with more general
cut-and-join operators $\S_{\l}$ on more general trace structures.
When writing partitions $\l$ for cut-and-join operators we omit parts
of length 1 so that in this example $\l = [2,1^{k-2}]\to [2]$. This
omission of length-1 parts corresponds to only considering skeleton
graphs, as discussed above.\footnote{In \cite{1002.2099} it was useful
  to think of the correlation function \eqref{eq:corrdelta} as an
  inner product of $\tr(\a_\n\;Z^{\dagger \otimes k})$ with the result
  of $\O_k$ acting on $\tr(\a_\m \;Z^{\otimes k})$.  The bra-ket
  notation of the inner product in \cite{1002.2099} is the same as the
  delta function here $\braket{\n}{\m} = \frac{k!}{|[\m]|\,|[\n]|}
  \delta(\S_\m \S_\n) = |\Sym([\m])| \delta_{\m=\n}$.}

Next we turn to the character expansion.  The symmetric group delta
function in \eqref{eq:corrdelta} can be expanded as a sum over
representations $R$ of $S_k$, also indexed by partitions, in terms of
the characters $\chi_R(\s)$ of $\s$ in the representations $R$ and the
$S_k$ representation dimensions $d_R = \chi_R(\id)$.  The identity is
$\delta(\s)= \frac{1}{k!} \sum_{R\vdash k} d_R \chi_R(\s)$.  Using the
fact that $d_R\chi_R(\S_\m\s)= \chi_R(\s) \chi_R(\S_\m)$, since
$\S_\m$ is central in $S_k$, and the relation between $\O_k$ and the
unitary group $U(N)$ dimension of the partition $R$, $N^k \chi_R(\O_k)
= k! \dim_N R$,\footnote{Characters of the cut-and-join operators
  $\chi_R(\S_\m)$ roughly correspond to $U(N)$ Casimirs, cf. Section
  2.7 of \cite{1002.2099} and original references in
  \cite{9407114,9411210}.}  the correlation function
\eqref{eq:corrdelta} has a character expansion
\begin{equation}
  \corr{ \tr(\a_\n \;Z^{\dagger\otimes k}) \;\; \tr(\a_\m \;Z^{\otimes k})} = \sum_{R \vdash k}  \frac{k!\;\chi_R (\a_\m)\chi_R (\a_\n)\dim_N R}{d_R} \label{eq:corrchar}
\end{equation}
This identity can also be seen by expanding the traces in terms of
$U(N)$ characters $\tr(\a_\m \;Z^{\otimes k}) =\sum_{R\vdash k}
\chi_R(\a_\m) \chi_R(Z)$ where the $U(N)$ character $\chi_R(Z)$ is itself
expanded in terms of traces $\chi_R(Z) = \frac{1}{k!} \sum_{\a \in
  S_k} \chi_R(\a) \tr(\a\;Z^{\otimes k})$.  The two-point function
of these characters $\chi_R(Z)$ is diagonal \cite{0111222}.  Note that
the dimension $\dim_N R$ vanishes if the number of parts $l(R)$ of the
partition $R$ exceeds $N$.

For general $S$ as in \eqref{eq:expansion1} the correlation functions
\eqref{eq:corrdelta} and \eqref{eq:corrchar} become
\begin{align}
  \corr{ \tr(\a_\n \;Z^{\dagger\otimes k}) \;\; \tr(\a_\m \;Z^{\otimes k})} & = \sum_{R \vdash k} \sum_{\l\vdash k}|[\l]| \tr(\a_\l \; S^{\otimes k}) \frac{\chi_R (\a_\m)\chi_R (\a_\n) \chi_R(\a_\l) }{d_R}  \label{eq:Zgraphexp} \\
& = \sum_{R \vdash k}   \frac{k!\;\chi_R (\a_\m)\chi_R (\a_\n)\chi_R(S)}{d_R}
\end{align}

Now we are in a position to expand the full $Z$ partition function.
It is useful to collect the expansions of the exponentials into
multi-trace operators indexed by partitions $\m$ of $k$, so that the
total number of $Z$ fields in each term is $k$.  Using  the Kontsevich-Miwa transformation \eqref{eq:KMid}
\begin{equation}
  e^{\sum_{k=1}^\infty t_k\tr( Z^k)} = \sum_{k=0}^\infty \sum_{\m \vdash k}\prod_{p=1}^k \frac{\big[t_p \tr(Z^p)\big]^{i_p(\m)}}{i_p(\m)!} = \sum_{k=0}^\infty \sum_{\m \vdash k}\prod_{p=1}^k \frac{\big[\tr(A^{-p})\tr(Z^p)\big]^{i_p(\m)}}{k^{i_p(\m)}i_p(\m)!}
\end{equation}
The denominator factors give the size of the conjugacy class from
\eqref{eq:factors} so
\begin{equation}
  e^{\sum_{k=1}^\infty t_k\tr( Z^k)} =  \sum_{k=0}^\infty \sum_{\m \vdash k}\frac{|[\m]|}{k!}\tr(\a_\m\;(A^{-1})^{\otimes k})\;\tr(\a_\m \;Z^{\otimes k}) =  \sum_{k=0}^\infty\sum_{R \vdash k} \chi_R(A^{-1}) \chi_R(Z) \label{eq:expexpansion}
\end{equation}
The $Z$ model partition function \eqref{eq:expansion1} becomes
\begin{align}
  \cZ(\{t\},\{{\ov t}\})  & =   \corr{e^{\sum_{k=1}^\infty {\ov t}_k \tr(Z^{\dagger k})}\;\;e^{\sum_{k=1}^\infty t_k\tr( Z^k)}} \nn \\
 & = \sum_{k=0}^\infty \sum_{\m,\n \vdash k} \frac{|[\m]|}{k!}\tr(\a_\m\;(A^{-1})^{\otimes k})\; \frac{|[\n]|}{k!}\tr(\a_\n\;(B^{-1})^{\otimes k} ) \; \corr{ \tr(\a_\n \;Z^{\dagger\otimes k}) \;\; \tr(\a_\m \;Z^{\otimes k})} \nn \\
& = \sum_{k=0}^\infty \sum_{R \vdash k} \frac{k!\;\chi_R(A^{-1}) \chi_R(B^{-1}) \chi_R(S) }{d_R} \label{eq:Zchar}
\end{align}
Noting that $\chi_R(S)$ for an $N\times N$ matrix $S$ vanishes if
$l(R) >N$, and since $n\geq N$ this constraint takes precedence over
those coming from $\chi_R(A^{-1})$ and $\chi_R(B^{-1})$, this concludes the proof of \eqref{eq:expansion2}.

\subsection{Character expansion of $F$ model}\label{sec:proofcharex}

A generalisation of the Itzykson-Zuber integral \cite{Itzykson:1979fi}
(see for example \cite{1005.5715}) can be used for the following
integral for the $F$ model with plain quadratic term
\eqref{eq:FFplain}
\begin{equation}
\corr{ \chi_R(A^{-1}FB^{-1}F^\dagger )}_{\textrm{plain}} =  \int [dF]^{\C}_{n\times n}\;\; e^{-\tr (FF^\dagger)}\chi_R(A^{-1}FB^{-1}F^\dagger) = \frac{|R|!\; \chi_R(A^{-1}) \chi_R(B^{-1})}{d_R} \label{eq:IZcomplex}
\end{equation}
For an $F$ model correlation function the vertices are
dictated by a partition $\l$ of $k$, so that the $i_p(\l)$ parts of $\l$ of
length $p$ correspond to $i_p(\l)$ $2p$-valent vertices.  Expanding with the formula \eqref{eq:IZcomplex}
\begin{align}
  \Big\langle \prod_{p=1}^k \left[\tr(A^{-1}FB^{-1}F^\dagger)^p\right]^{i_p(\l)}\Big\rangle_{\textrm{plain}} =&  \corr{\tr( \a_\l\; (A^{-1}FB^{-1}F^\dagger)^{\otimes k})}_{\textrm{plain}} \nn \\
=&  \sum_{R\vdash k}  \frac{k!\chi_R(\a_\l) \chi_R(A^{-1}) \chi_R(B^{-1})}{d_R} 
\end{align}
Expanding out the characters $\chi_R(A^{-1})$ and $\chi_R(B^{-1})$ as
multi-traces gives
\begin{equation}
\sum_{R\vdash k} \sum_{\m,\n\vdash k} \frac{|[\n]|\,|[\m]|}{k!} \tr(\a_\m\;(A^{-1})^{\otimes k})  \tr(\a_\n\;(B^{-1})^{\otimes k}) \frac{\chi_R(\a_\l)\chi_R(\a_\m) \chi_R(\a_\n)}{d_R} \label{eq:Fgrapgexp}
\end{equation}
This expansion is important because we see how the faces of the
$F$ model diagrams are controlled by the partitions $\m$ and $\n$ that
organise the couplings $\{t\}$ and $\{{\ov t}\}$.  Comparing
\eqref{eq:Fgrapgexp} to its $Z$ model equivalent \eqref{eq:Zgraphexp}
and picking out the terms for particular face configurations in the respective models
\begin{align}
  k! \corr{\tr( \S_\l\; (A^{-1}FB^{-1}F^\dagger)^{\otimes p})}_{\textrm{plain, }\m,\n\textrm{ term}}&  =    \corr{ \tr(\S_\n \;Z^{\dagger\otimes k}) \;\; \tr(\S_\m \;Z^{\otimes k})}_{\l\textrm{ term}} \nn \\
& = \sum_{R\vdash k}\frac{\chi_R(\S_\n)\chi_R(\S_\l) \chi_R(\S_\m)}{d_R} \nn \\
& = k!\; \delta(\S_\n \S_{\l}\S_{\m})
\end{align}
In other words both models are doing nothing other than computing the conjugacy class algebra of the symmetric group.

If we use the corrected propagator \eqref{eq:FFprop} then there are no
2-valent vertices in the $F$ model, corresponding to the fact that all
parallel propagators of the $Z$ model are bunched into single edges.
Thus with the corrected propagator for the $F$ model the partitions
$\l$ of vertices have no parts of length 1, $i_1(\l) = 0$.  To relate
this to the above analysis above define $\l^r = \l + [1^{r}]$ so that
\begin{align}
&  k! \corr{\tr(\S_\l \;(A^{-1}F B^{-1}F^\dagger)^{\otimes k})  } \;\;  = \;\; k!\sum_{r=0}^\infty  \corr{\tr(\S_{\l^r} \;(A^{-1}FB^{-1}F^\dagger )^{\otimes k+r})  }_{\textrm{plain}} \nn\\
&\quad  = \sum_{r=0}^\infty  \sum_{\m,\n\vdash k+r} \tr(\a_\m\;(A^{-1})^{\otimes k+r})\;\tr(\a_\n\;(B^{-1})^{\otimes k+r})\;  \corr{  \tr(\S_\n \;Z^{\dagger\otimes k+r}) \;\;\tr(\S_\m \;Z^{\otimes k+r})}_{\S_{\l^k} \textrm{ term}} \nn
\end{align}
Even if we fix the structure of $\m,\n,\l$ there are
still an infinite number of $Z$ diagrams for every $F$ diagram because
in the $Z$ model we can have any number of bunched parallel
propagators at each edge, corresponding to adding arbitrary numbers of
2-valent vertices to the plain $F$ model.

In the original vanilla $Z$ model with $S = \bI_N$ picking a
particular power of $N$ in the expansion of the correlation function
\eqref{eq:corrdelta} corresponds to fixing the genus of the
diagrams in the $Z$ model expansion.  Looking at the expansion of
$\O_k$ in terms of cut-and-join operators $\S_\l$ in equation
\eqref{eq:Onexp}, fixing the genus means we only consider $\l$
composed from a fixed number of transpositions $T([\l]) = p$.  The
generic $\l$ with this property is $[2^p]$, followed by cases where
the transpositions blend into longer cycles
$[3,2^{p-2}],[4,2^{p-3}],[3,3,2^{p-4}]$ etc.  For example $[2,2] \to
[3]$ corresponds to the degenerate multiplication of two
transpositions when they share an element $i$ to get a 3-cycle
$(ik)(ij) = (ijk)$.

In the $F$ model $\l = [2^p]$ means $p$ 4-valent vertices in the
graph.  The other terms for fixed genus, such as $[3,2^{p-2}]$,
correspond to vertices in the $F$ model colliding to create higher
valency vertices.  For example the collision of two four-valent
vertices into a single six-valency vertex for the dual to the $Z$
model torus two-point function in Figure \ref{fig:oomm-tor} is the
counterpart of $[2,2] \to [3]$.  The limit with only the generic
cut-and-join operator $\S_{[2^p]}$, corresponding to having only
4-valent vertices in the $F$ model, is discussed in Section
\ref{sec:klimit}.

Note that the number of edges $E$ in the $Z$ or $F$ model diagram is
given by a sum over the faces/vertices weighted by their valency $2E =
\sum_{p=1}^k 2p\,i_p(\l) = \sum_{p=1} p\, (i_p(\m)+i_p(\n))$.  This is
automatic since $\l$, $\m$ and $\n$ are each partitions of $E=k$.

Finally, to prove equation \eqref{eq:expansion3} for the full
expansion of the $F$ model partition function expand the exponential
using \eqref{eq:expexpansion}. With
couplings $s_k = \frac{1}{k} \tr(S^k)$  to the $2k$-valent vertices
\begin{align}
&  \int [dF]^{\C}_{n\times n} \;\;e^{-\tr (FF^\dagger) +\sum_{k=1}^\infty s_k \tr(A^{-1}FB^{-1}F^\dagger)^k} \nn \\
  & = \sum_{k=0}^\infty \sum_{\l \vdash k} \frac{|[\l]|}{k!}\tr(\a_\l\;S^{\otimes k})\;\corr{ \tr(\a_\l \;(A^{-1}FB^{-1}F^\dagger)^{\otimes k})}_{\textrm{plain}} \nn \\
&= \sum_{k=0}^\infty  \sum_{R \vdash k} \frac{|R|!\;\chi_R(S) \chi_R(A^{-1}) \chi_R(B^{-1})}{d_R} \label{eq:Fcharexp}
\end{align}

\subsection{A large $k$ limit}\label{sec:klimit}

It was shown in \cite{0209215,1002.2099} that if you take a given
correlation function in the $Z$ model and fix the genus, corresponding
to a particular power of $N$, then in the large $k$ limit of operators
with many powers of $Z$ the cut-and-join operator $\S_{[2^p]}$
dominates (cf. the torus two-point function in \eqref{eq:2pttorus})
\begin{align}
  \corr{\tr(\a_\n \;Z^{\dagger\otimes k})\;\;\tr(\a_\m \;Z^{\otimes k}) }_{N^{k-p} \textrm{ term}} &  = \frac{k! N^k}{|[\m]|\,|[\n]|} \sum_{\l: T([\l]) = p} \frac{1}{N^{p}} \; \delta(\S_\n \S_\l \S_\m) \nn \\
& \stackrel{k \to \infty}{\to} \frac{k! N^{k-p}}{|[\m]|\,|[\n]|} \; \delta(\S_\n \S_{[2^p]} \S_\m) 
\end{align}
In other words, if we take the expansion of $\O_k$ in cut-and-join
operators in equation \eqref{eq:Onexp} and take this limit for each
inverse power of $N$ then the $\S_{[2^p]}$ term always dominates as a function of $k$ in the correlation functions at this genus
\begin{equation}
  \O_k = \sum_{p=0}^{k} \frac{1}{N^p}\sum_{\l : T(\l) =p} \; \S_\l\quad \stackrel{k \to \infty}{\to} \quad\sum_{p=0}^{\infty} \frac{1}{N^p}\;\S_{[2^p]} 
\end{equation}
For a fixed genus the terms we lose in this limit correspond to the
degenerate cases where transpositions collide to give higher cycles.
These degenerate cases appear with less frequency so they are
suppressed for large $k$.

Since in this large $k$ limit $\S_{[2^p]} \sim \frac{1}{p!}
\S_{[2]}^p $, $\O_k$ can be said to exponentiate $\O_k \to \exp
\left(\frac{1}{N} \S_{[2]}\right)$.  The sum over
transpositions $\S_{[2]}$ has a simple interpretation in terms of
either splitting or joining a trace.  Geometrically this means
that $n$ punctured genus $g$ surfaces factorise into 3-punctured
spheres.  See \cite{1002.2099} for more details.

Here $N$ has been treated as a book-keeping device for the genus.  If
we allow $N$ to be a number, then the limit must be taken delicately.
Really we are taking the double-scaling BMN limit $k \sim N^{\ha} \to
\infty$ \cite{0202021} and expanding with the non-planar coupling $g_2
= \frac{k^2}{N} < 1$, cf. \cite{1002.2099}.  If $k$ grows faster than
$N^{\ha}$ then other terms become significant and spoil the simple
exponentiation $\O_k \to \exp \left(\frac{1}{N} \S_{[2]}\right)$.

In terms of the $F$ model, restricting to cut-and-join operators of
the form $\S_{[2^p]}$ means that we only have 4-vertices in the model.
In terms of the couplings $s_k$, we are setting $s_k = 0$ for $k \geq
3$ and only retaining 4-vertices $s_1=N, s_2 = \frac{N}{2}$.  Diagrams
involving higher-valency vertices resulting from collisions of
4-vertices are dropped.

In the 2d topological gravity case the Hermitian matrix model with couplings
$ \sum_{k=1}t_k \tr(M^k)$ can also be rearranged exactly into a
Kontsevich-Penner model \emph{before} taking the double-scaling limit. This
is the Chekhov-Makeenko model \cite{9202006} and is the
Hermitian version of the complex matrix model duality discussed
here. The dual model has vertices of all valency, including odd ones.
Taking the double-scaling limit eliminates all but the 3-valent terms,
giving the standard Kontsevich model.

\subsection{Hurwitz theory for $\CP^1 \setminus \{0,\infty\}$}\label{sec:Hurwitz}

The formulae for the correlation functions as sums over cut-and-join
operators \eqref{eq:corrdelta} can be interpreted in terms of
Hurwitz numbers that count holomorphic maps from the Riemann surface
$\mathbb{S}_g$ on which a graph is drawn to the sphere $\CP^1$
with three branch points.  For a $k$-sheeted covering of $\CP^1$ by a
genus $g$ surface with ramification profiles $\m,\n$ and
$\l$ the number of coverings is
\begin{equation}
  \textrm{Cov}_{k}^g (\n,\l,\m)  :=
  \sum_{f(\n,\l,\m):\, \mathbb{S}_g \to \CP^1} \frac{1}{|\Aut(f)|}\;\;
 = \;\;\frac{1}{k!} \delta(\S_{\n} \S_{\l}\S_{\m})
\end{equation}
The maps are counted up to automorphisms of the covering map $f :
\mathbb{S}_g \to \CP^1$.  The genus $g$ for which this is
non-vanishing is given by the Riemann-Hurwitz theorem, which relates
the genus $g$ of $\mathbb{S}_g$ to the branching numbers at each
branch point
\begin{equation}
  2g-2 = -2k+ T([\n])+ T([\l]) + T([\m]) \label{eq:RH}
\end{equation}
The branching number at each branch point is given by the minimum
number of transpositions needed to build the conjugacy class
corresponding to the partition.\footnote{There is a review of Hurwitz
  theory in \cite{9411210}. The relation of these cut-and-join
  operators and Hurwitz theory to integrable hierarchies is summarised
  in \cite{0904.4227}.}  For our case this formula is just a restatement of the Euler characteristic formula \eqref{eq:Eulerchar}.

So each correlation function is a sum over Hurwitz numbers for maps to $\CP^1$ with three branch points
\begin{equation}
\boxed{  \corr{ \tr(\S_\n \;Z^{\dagger \otimes k}) \;\; \tr(\S_\m \;Z^{\otimes k})}  =(k!)^2\sum_{\l \vdash k }N^{C([\l])}\;  \textrm{Cov}_{k} (\n,\l,\m) }
\end{equation}

There is a similar story for the Hermitian matrix model
\cite{9212108,1002.1634}.  The relation with the complex matrix model
discussed here is simple: just replace one of the profiles $\n$ with
$[2^{\frac{k}{2}}]$ to account for the Hermitian matrix model
propagator.  There is then an identity that comes from setting ${\ov
  t}_k = \delta_{k2}$, known in the literature \cite{9212108,1005.5715},
\begin{align}
  \cZ(\{t\},\{s\}) & = \int [dM]^H_{N\times N} \;\; e^{-\tr(S^{-1}MS^{-1}M) + \sum_{k=1}^{\infty} t_k \tr(M^k)}\nn\\
& = \int [dZ]^{\C}_{N\times N}\;\; e^{-\tr(S^{-\frac{1}{2}}ZS^{-\frac{1}{2}}Z^\dagger) + \sum_{k=1}^\infty t_k \tr(Z^k) +\tr( Z^{\dagger 2})} \nn \\
& = \int [dF]^{\C}_{N\times N} \;\;e^{-\tr FF^\dagger + \tr(A^{-1}FSF^\dagger)^2}
\end{align}
In the expansion of the Hermitian model in Feynman diagrams,
$k$-valent faces come with a coupling $s_k$ while $k$-valent vertices
come with a coupling $t_k$ \cite{9212108}.  Graph duality in this
model  \cite{Gopakumar} just exchanges $\{t\}$ for $\{s\}$.  The
identity with the $Z$ model in the second line can be seen directly
since the $\tr(Z^{\dagger 2})$ term allows each propagator to loop
back to the holomorphic operator built out of $Z$'s, which is then
equivalent to the Hermitian matrix model operator in the Hermitian
correlation function.  The difference in powers of $S$ in the actions
arises because there are now two edges in the $Z$ model for every edge
in the Hermitian model.  The final line follows from the symmetry in
the character expansion \eqref{eq:expansion2}; the identity between
the first and third line was proved in \cite{9212108} using the same
techniques as used in Section \ref{sec:proof}.\footnote{Holomorphic maps
  onto $\CP^1$ with just three branch points are special.
  Belyi's theorem \cite{Belyi} states that a non-singular Riemann
  surface is an algebraic curve defined over the algebraic numbers
  ${\ov \Q}$ if and only if there is a holomorphic map of the Riemann
  surface onto $\CP^1$ with only three branch points.  For the
  Hermitian matrix model, where all the ramification orders over one
  of the branch points are 2, the Belyi map is a special type called
  `clean' \cite{1002.1634}.}

Taking the limit described in Section \ref{sec:klimit} so that
$\S_{\l} \to \S_{[2^p]} \sim \frac{1}{p!} \S_{[2]}^p$, which for the
$F$ model means restricting to only 4-valent vertices, the Hurwitz
numbers become \emph{double Hurwitz numbers}.  Double Hurwitz numbers
have two fixed branching profiles $\m,\n$ for branch points at $0$ and
$\infty$ and a remaining arbitrary number of simple branch points
(with profile $[2]$).\footnote{Single Hurwitz numbers occur when there is no branching at the second point, i.e. $\n = [1^k]$.  The generating function for single Hurwitz numbers is obtained from the $Z$ model with ${\ov t}_{k} = \delta_{k1}$.}  In this limit the partition function becomes
the $\tau$-function for the Toda lattice hierarchy that appears in the
Gromov-Witten theory of $\CP^1$
\cite{math/9912166,math.AG/0004128}\footnote{The map to the variables
  in \cite{math.AG/0004128} is $p_\m = \tr(\a_\m\;(A^{-1})^{\otimes k}) $, $p_\n' = \tr(\a_\n\;(B^{-1})^{\otimes k})$, $q = s_1$ and $\b = \frac{2s_2}{s_1^2}$.}
\begin{align}
  \cZ(\{t\},\{{\ov t}\},\{s\}) & \to \sum_{k=0}^\infty \sum_{\m,\n\vdash k} \sum_{p=0}^\infty  \frac{s_1^{k-2p}(2s_2)^p}{k!p!}\tr(\a_\m\;(A^{-1})^{\otimes k})\;\tr(\a_\n\;(B^{-1})^{\otimes k} ) \; \delta(\S_\m \S_{[2]}^p \S_\n)  \nn \\
& = \sum_{k=0}^\infty \sum_{\m,\n\vdash k} \sum_{p=0}^\infty  \frac{s_1^{k-2p}(2s_2)^p}{p!}\tr(\a_\m\;(A^{-1})^{\otimes k})\;\tr(\a_\n\;(B^{-1})^{\otimes k} ) \;\textrm{Cov}_k(\m,\n,\underbrace{[2], \cdots [2]}_{p \textrm{ times}})  \nn 
\end{align}

It should be possible to make this picture more precise: the $Z$ and
$F$ models compute relative Gromov-Witten invariants for the
topological $A$ model on $\CP^1$ with two points marked at $0$ and
$\infty$, cf. \cite{math/0204305}.  The holes in the $F$ worldsheet
corresponding to holomorphic $Z$ operators wrap around $0$ while the
holes corresponding to antiholomorphic $Z^\dagger$ operators wrap
around $\infty$.  The parts of the cut-and-join operator $\S_{\l}$
then map to gravitational descendants $\tau_{\l_i}(w)$ of the K\"ahler
class $w$.  The full details of this correspondence require the
technology of completed cycles \cite{math/0204305}.

The appearance of $\CP^1\setminus\{0,\infty\}$ is curious because of
its appearance both as an auxilliary curve for the normal matrix model
description of the $c=1$ string \cite{0302106} and in the topological
B model set-up for the $c=1$ string in \cite{0312085}.  The free
energy of the $c=1$ string at the self-dual radius is also known to
agree with that for the topological A model on the conifold resolved
by a sphere in the limit where the complexified K\"ahler class vanishes $t\to 0$ \cite{0205297}.

\section{The $F$ model and the cell decomposition of $\cM_{g,n}$}\label{sec:discussion}

The (Deligne-Mumford-compactified) moduli space $\cM_{g,n}$ of Riemann
surfaces of genus $g$ punctured $n$ times can be extended to the space
of `ribbon' or `fat' graphs by replacing each puncture with a boundary
of length $\ell_i$ for $i = 1, \dots n$.  This extended
\emph{decorated} moduli space $\cM_{g,n} \times \R_+^n$ has real
dimension $6g-6+3n$.  Due to the work of Mumford, Thurston, Strebel
\cite{strebel}, Harer \cite{harer}, Penner \cite{Penner} and others,
any point in this moduli space can be obtained by considering
connected graphs with lengths assigned to each edge.

The Penner Hermitian matrix model \cite{Penner}
\begin{equation}
  \int [dQ]^H\;\; e^{\tr(Q)+\tr\log(1-Q)}  = \int [dQ]^H\;\; e^{-\sum_{k=2}^\infty \frac{1}{k}\tr(Q^k)}
\end{equation}
gives a cell decomposition of $\cM_{g,n} \times \R_+^n$.  Each graph
of the Penner model with $n$ faces and genus $g$ corresponds to one of
the cells.  The top-dimensional cells in $\cM_{g,n} \times \R_+^n$ are
swept out by the lengths of the $6g-6+3n$ edges of the Feynman graphs
with only 3-valent vertices.  Lower-dimensional cells in the cell
decomposition arise when we shrink an edge, colliding two
vertices into a higher-valency vertex.  Because of the extra factor of
$-1$ from the vertices with each lowering of dimension, the Penner
model calculates the virtual Euler characteristic of $\cM_{g,n}$.  The
symmetry factors of the Feynman graphs account for the fact that
$\cM_{g,n}$ is an orbifold space.

Konstevich \cite{Kontsevich:1992ti} adapted the 3-valent version of
this model to give a generating function for the correlators of 2d
topological gravity, which calculate intersection numbers on
$\cM_{g,n}$ \cite{Witten:1989ig}.  The couplings $t_n$ to the
operators are encoded in a matrix $Z$ by the transformation $t_k =
\frac{1}{k} \tr(Z^{-k})$.  This constant matrix modifies the quadratic
term in the Kontsevich matrix model
\begin{equation}
  \int [dM]^H\;\; e^{\tr(-\frac{1}{2}ZM^2 + \frac{1}{6} M^3)} \label{eq:kont}
\end{equation}
In the expansion of this partition function, each Feynman graph with
$n$ faces and genus $g$ can be written as an integral over the
corresponding top-dimensional cell in $\cM_{g,n} \times \R_+^n$ using
the Schwinger parameterisation of the propagators
\begin{equation}
 \corr{M^i_j\; M^k_l} = \delta^i_l \delta^k_j \frac{2}{z_i + z_j} = 2\delta^i_l \delta^k_j\;\int_0^\infty dp_e\; e^{-p_e(z_i+z_j)} \label{eq:kontprop}
\end{equation}
By integrating over the $6g-6+3n$ lengths $p_e$ of the edges of the
graph the whole of the cell is covered.  Separating out the integral
over the boundary lengths $\R_+^n$ (which correspond to sums of the
edges around each boundary) one is left with an integral over
$\cM_{g,n}$ corresponding to the closed string correlation function
\cite{Kontsevich:1992ti} (see \cite{9201003,0504229} for concise
summaries).

The $F$-model only has vertices of even valency, which suggests that
it localises on lower dimensional cells in the complex for $\cM_{g,n}
\times \R_+^n$.  The maximal-dimensional cell for the $F$ model,
corresponding to all vertices of valency 4, has dimension $4g-4+2n$
corresponding to the number of edges in the diagrams.  This
localisation on degenerate subspaces of the moduli space was noticed in
\cite{0602226,0703141} when considering extremal 4-point functions in
$\cN=4$ SYM.

Following the example of the Kontsevich-Penner Hermitian model of
Chekhov and Makeenko \cite{9202006} we will write each generic $F$
model graph as an integral over a discrete version of a
lower-dimensional cell in $\cM_{g,n}\times \R_+^n$ \cite{9302014}.  In
the continuum limit the Chekhov-Makeenko model reduces to the
Kontsevich model.  In the continuum limit of the $F$ model we get a
4-valent model which is the complex analogue of the Kontsevich model.

With discretisation parameter $\ve$ rewrite $A = \sqrt{N} e^{\ve L}$
and $B = \sqrt{N} e^{\ve M}$, i.e. $a_i = \sqrt{N} e^{\ve l_i}$ and
$b_j = \sqrt{N} e^{\ve m_j}$.  The propagator \eqref{eq:Fscaledprop}
of the rescaled $F$ model \eqref{eq:Fscaled} can then be written
\begin{equation}
  \frac{1}{a_ib_j - N} = \frac{1}{Ne^{\ve l_i}e^{\ve m_j} - N} = \frac{1}{N}\sum_{p=1}^\infty e^{-p\ve( l_i+ m_j)} \label{eq:disc}
\end{equation}
The sum is a discrete Schwinger parameterisation of the edge length
for the propagator.  Each summand comes from an edge of the $F$ graph
with integer length $p\ve$, corresponding to a different number $p$ of
bunched propagators in the dual $Z$ model graph,
cf. \eqref{eq:bigsum}.\footnote{Such a discrete metric on the moduli
  space also appears for the Hermitian matrix model in
  \cite{0803.2681,0911.0658}, where the integer lengths also
  correspond to the number of bunched propagators between the
  vertices.}  The integer length $\ell$ of the boundary of each face
is the valency of the vertex of the dual $Z$ model graph, or in other
words the power of the operator $\tr(Z^{\ell})$ or $\tr(Z^{\dagger \ell})$.

For an $F$ model graph with $V$ vertices, $E$ edges and faces
corresponding to $Z$ vertices labelled by $f$ (and colour index $i_f$)
and $Z^\dagger$ faces labelled by $g$ (and colour index $j_g$) the
contribution is
\begin{align}
  cN^V  \sum_{\{i_f\},\{j_g\}}\prod_{ E \textrm{ edges}} \frac{1}{a_{i_f}b_{j_g} - N} & =   cN^{V-E}  \sum_{\{i_f\},\{j_g\}}\prod_{ E \textrm{ edges}} \frac{1}{e^{\ve l_{i_f}}e^{\ve m_{j_g}} - 1} \nn \\
& =   cN^{V-E} \sum_{\{i_f\},\{j_g\}} \prod_{r=1}^E \sum_{p_r = 1}^\infty e^{-\ve p_r (l_{i_f}+m_{j_g})} \label{eq:discsum}
\end{align}
$c$ is the symmetry factor for the graph.  The discrete sums for each
of its $E$ propagators give a sum over discrete points in an
$E$-dimensional cell of $\cM_{g,n}\times \R_+^n$.  Each point
corresponds to a different $Z$ model graph with different numbers of
bunched propagators between each set of vertices, given by the
integers $p_r$.  A given $Z$ model correlation function gets
contributions from a finite number of graphs, so the closed string
correlation function must localise on only a finite number of points
in the moduli space.

The propagator in its discrete Schwinger parametrisation \eqref{eq:disc} has a continuum limit as $\ve \to 0$
\begin{equation}
   \lim_{\ve \to 0} \sum_{p=1}^\infty \ve e^{-p\ve(l_i+m_j)} = \int_{0}^\infty dp e^{-p(l_i+m_j)}  = \frac{1}{l_i + m_j} \label{eq:Gpropagator}
\end{equation}
The model with this propagator arises in a double-scaling limit of the
$F$ model where we take $N$ large $N \sim \frac{1}{\ve^2}$.  Rescaling
the $F$ model matrix $F = \sqrt{\e}G$ the action \eqref{eq:Fscaled} becomes
\begin{align}
  &  -\tr(AFBF^\dagger) + N\textstyle\sum_{k=1}^\infty \tfrac{1}{k} \tr(FF^\dagger)^k \nn \\
& = - N\ve\tr\left[ (1+\ve L + \cO(\ve^2) )\;G\;(1+\ve M + \cO(\ve^2) )\;G^\dagger\right]+ N \tr\left[ \ve GG^\dagger +\ve^2\tfrac{1}{2}GG^\dagger GG^\dagger + \cO(\ve^3)\right] \nn \\
&  = - N\ve^2 \tr(G^\dagger LG) -N \ve^2 \tr(GMG^\dagger) + N\ve^2\tfrac{1}{2} \tr(GG^\dagger GG^\dagger) + N\cO(\ve^3) \nn \\
& \to - \tr(G^\dagger LG) -\tr(GMG^\dagger) + \tfrac{1}{2}\tr(GG^\dagger GG^\dagger)
\end{align}
This $G$ model is a complex analogue of the Kontsevich model \eqref{eq:kont}
\begin{equation}
  \int [dG]^\C_{N \times N} \;\;e^{-\tr(G^\dagger LG) -\tr(GMG^\dagger) + \tfrac{1}{2}\tr(GG^\dagger GG^\dagger)}
\end{equation}
It has the propagator identified in \eqref{eq:Gpropagator} which is similar to that of the Kontsevich model \eqref{eq:kontprop}
\begin{equation}
  \corr{ G^i_j\; G^\dagger{}^k_l} = \delta^i_l\delta^k_j\;\frac{1}{l_i + m_j}
\end{equation}

Looking at an individual Feynman graph of the $F$ model we can also
see that only 4-valent graphs survive in this double-scaling limit.
Suppose a graph has $i_{k}$ vertices of even valency $2k$. There are
$V = \sum_{k=2}^\infty i_{k}$ vertices in total and $E = \frac{1}{2}
\sum_{k=2}^\infty 2k i_{k} =\sum_{k=2}^\infty k i_{k} $ edges.
Setting $N = \frac{1}{\ve^2}$ and taking the limit $\ve \to 0$ then
the expression \eqref{eq:discsum} is only non-vanishing if there is a
factor of $\ve$ for each edge (cf. \eqref{eq:Gpropagator})
\begin{equation}
  N^{V-E} = \ve^E
\end{equation}
This means that $E = 2V$, which is satisfied if only $i_2$ is non-zero.  In this case we get
\begin{align}
\lim_{\ve \to 0} c  \sum_{\{i_f\},\{j_g\}} \prod_{r=1}^E \left[\ve\sum_{p_r = 1}^\infty e^{-\ve p_r (l_{i_f}+m_{j_g})}\right] & = c\sum_{\{i_f\},\{j_g\}} \prod_{r=1}^E \int_{0}^\infty d p_r e^{- p_r (l_{i_f}+m_{j_g})}\label{eq:integra}  \\
& = c\sum_{\{i_f\},\{j_g\}}\prod_{ E \textrm{ edges}} \frac{1}{l_{i_f}+m_{j_g}} 
\end{align}
This comes from the corresponding graph of the $G$ model.

The integrals over the worldsheet boundary lengths $\R_+^n$ must be
decoupled with care, since there is at least one relationship between
the boundary lengths: the sum of the $Z$ boundary lengths must equal
the sum of the $Z^\dagger$ boundary lengths.  For example in Figure
\ref{fig:oomm-sph-3} the three-point function has only two
independent boundary lengths, not three.  Once this is done, it is
still not clear what quantity we are integrating over (a subspace of)
$\cM_{g,n}$.

The fact that the $G$ model has only 4-valent vertices relates it
to the BMN limit from Section \ref{sec:klimit}, which arose
from limiting the $F$ model to 4-valent vertices. However it is
still not clear what the $G$ model is calculating in this
context.  The obvious link would be if in the continuum limiting
process the discrete sum only got contributions when $p
\sim\frac{1}{\ve} = \sqrt{N}$ (BMN-length operators propagating
between vertices of the $Z$ model), but this is not the case.  This
issue is left for the future.

\section{Conclusions and future directions}

In this paper we have studied correlation functions that correspond
both to tachyon scattering for the $c=1,R=1$ non-critical string and
to a half-BPS sector in free 4d $\cN=4$ super-Yang-Mills.  In the $Z$
complex matrix model the closed string insertions correspond to
\emph{vertices} of the Feynman diagrams.  The $Z$ model is precisely
dual to another complex matrix model called the $F$ model.  In the $F$
model the closed string insertions are now associated to the
\emph{faces} of its Feynman diagrams.  This duality can be shown using
character expansions, or by integrating in and out fields to see the
graph duality dynamically, following the programme set out by Gopakumar
\cite{Gopakumar}.

Using the example of the Kontsevich model, the correlation functions
of the $F$ model can be written as sums over discrete points in
subspaces of the moduli space of punctured Riemann surfaces.  These
discrete points correspond to ribbon graphs with integer-length edges.

This complex matrix model duality could provide a prototype for
understanding the AdS/CFT duality microscopically.  It may be possible
to rewrite (perhaps just free) $\cN=4$ super Yang-Mills as a dual
theory, where local operators and interaction vertices from $\cN=4$
SYM correspond to faces of the dual Feynman graphs. The correlation
functions of this dual theory would be easier to write as string
moduli space integrals, following the Kontsevich schema. At non-zero
coupling, summing over the interaction vertex holes should remove the
D3 branes and alter the background to $AdS_5 \times S^5$.

There is a long way to go to realise this goal.  In contrast to the
programme set out in \cite{0308184,0402063,0504229}, here we have
dropped the spacetime dependence of the $\cN=4$ correlation functions
to focus on the combinatorial index structure from which the
non-planar expansion comes.\footnote{Another way of introducing the
  spacetime dependence is discussed in \cite{0003065}.}  Really the
sector of $\cN=4$ we have studied just computes a metric on
multi-trace half-BPS states and doesn't contain any spacetime
information.$\textsuperscript{\ref{foot:metric}}$ More general correlation
functions not only include more interesting spacetime dependence but
should also get contributions from the full moduli space of punctured
Riemann surfaces, not just subspaces.

As long as we keep the separation between holomorphic and
anti-holomorphic operators, it should be straightforward to introduce
other complex scalar fields into this duality.\footnote{See Appendix
  \ref{sec:2more} for a sketch of how this might work with two
  complex matrices.}  The non-planar expansion of the free theory in
terms of cut-and-join operators discussed in Section \ref{sec:halfbps}
follows through with little modification \cite{1002.2099}.  Allowing
the scalars to be real, or introducing fermions and the gauge boson,
would introduce more complication.

From the string side, it is important to understand how the reduction
to the $c=1$ string from the full IIB string on $AdS_5 \times S^5$
works.  Both contain a Liouville direction and it is tempting to
identify the $R=1$ limit with the small radius limit of the bulk
geometry corresponding to free SYM.  The $c=1$ string in this limit is
known to have various topological coset descriptions
\cite{Witten:1991yr,9301083,9302027}, so perhaps it is possible to use
the cohomological reduction techniques of \cite{1001.1344} for sigma
models with supersymmetric target spaces to reduce the full bulk coset.

On the other hand, an alternative strategy would be to take a
topological description of the $c=1,R=1$ string and try to include the
full $PU(2,2|4)$ symmetry of free 4d $\cN=4$ SYM.  For example, it is
known that the free energy of the $c=1,R=1$ string agrees with that
for the $t\to 0$ limit of the topological $A$ model on the resolved
conifold \cite{0205297}.  Once one understands how tachyon scattering
is reproduced in that setting, one can think about how to include more
of the spectrum of $\cN=4$.  An $A$ model approach seems promising
given that the correlation functions were shown in Section
\ref{sec:Hurwitz} to count holomorphic maps onto $\CP^1$. A tentative
connection to the $A$ model on $\CP^1$ was made at the end of that
Section.

Focusing now on the $F$ model there are several areas that merit
further study:
\begin{itemize}
\item A brane interpretation of the $F$-model would be welcome,
  perhaps along the lines of the relation between the Kontsevich model
  and the open string field theory of FZZT branes for 2d topological
  gravity derived by Gaiotto and Rastelli \cite{0312196}.  Such an
  interpretation of the $W_\infty$ model has been discussed in
  \cite{0406106,0512217}.
\item What mechanism localises the integral over the moduli space to
  discrete points in Section \ref{sec:discussion}?  Given that this is
  also a feature of the Hermitian matrix model
  \cite{9302014,0911.0658}, does it always arise in free theories?
\item The $Z$ model and the $F$ model capture tachyon scattering
  for the $c=1$ string at the self-dual radius, but they do not
  include all of the discrete states or the $SU(2)$ symmetry at this
  particular radius.  Perhaps vortices appear in the $F$ model as
  holomorphic and antiholomorphic operators $\tr(F^k)$ and
  $\tr(F^{\dagger k})$ like their appearance in the similar six-vertex
  model \cite{9911023}.

\item The Toda integrable hierarchy structure of the $c=1$ string at
  the self-dual radius has not been discussed here from the point of
  view of the $F$ model.
\item An algebraic geometry interpretation of the $F$ model might
  correspond to the limiting case discussed in \cite{Witten:1991mk} for the
  $c=1,R=1$ string.
\item The authors of \cite{0312085} reproduced both the Kontsevich
  model and the $c=1,R=1$ tachyonic scattering matrix by considering
  non-compact branes in the topological B model in a deformed conifold
  background.  A model similar to the $Z$ model was also studied in
  \cite{0411280} for the $c=1,R=1$ string.  Where does the $F$ model
  fit into this picture?
\end{itemize}

\begin{appendix}

\hspace{1cm}

{\bf Acknowledgements:} We thank Sasha Alexandrov, Robert de Mello
Koch, Rajesh Gopakumar, Vladimir Kazakov, Sanjaye Ramgoolam and Volker
Schomerus for discussions and comments on early drafts.  We are also
very grateful to the organisers of the Second Johannesburg Workshop on
String Theory, during which many of these ideas were aired.

\section{Examples for $F$ model}\label{sec:examples}

In this section we consider explicit examples of how the $F$ model
Feynman diagrams reproduce the $Z$ model correlation functions.  First
we consider the $F$ model expansion.

A single 4-valent vertex $\corr{ \tr(FF^\dagger FF^\dagger)}$ can only
contract with itself to form a planar graph that is dual to the planar
3-point function of the $Z$ model, cf. Figure
\ref{fig:oomm-sph-3} and Section \ref{sec:m03} for the full expansion.

For two 4-valent vertices $\corr{ \tr(FF^\dagger
  FF^\dagger)\tr(FF^\dagger FF^\dagger)}$ there are several possible
diagrams:
  \begin{enumerate}
  \item Each vertex entirely contracted with itself is disconnected
    and corresponds to two disconnected 3-point functions.
  \item Each vertex with just one self-contraction each corresponds to
    the lollipop planar $1\to 3$ 4-point function, see Section \ref{sec:m04ex}.
  \item The planar diagram with each propagator connecting the two vertices
    corresponds to the planar $2\to 2$ 4-point function all
    connected in a loop, see Section \ref{sec:m04non}.
  \item The torus diagram with each propagator connecting the two vertices corresponds
    to the torus 2-point function, see Section \ref{sec:m12}.
  \end{enumerate}

All the cases for a single 6-valent vertex $\corr{ \tr(FF^\dagger
  FF^\dagger FF^\dagger)}$ correspond to the collision of the two
4-valent vertices considered above:
\begin{enumerate}
\item One way of contracting it in a planar way with 4 faces
  corresponds to the $1\to 3$ 4-point function Y diagram, cf. Section
  \ref{sec:m04ex}, that arises from colliding the two vertices from case
  2. above.
\item A topologically distinct way of contracting it in a planar way
  corresponds to the $2\to 2$ degenerate $\corr{ \tr(Z^p)
    \tr(Z^{\dagger p+q}) \tr(Z^{\dagger q+r}) \tr(Z^r)}$ from the
  collision of vertices in 3. above, cf. Section \ref{sec:m04non}.
\item The vertex contracted with itself on the torus has two faces and is
  $\cM_{1,2}$, cf. Section \ref{sec:m12}.  This results from the collision of the vertices in 4. above.
\end{enumerate}

\subsection{$\cM_{0,3}$}\label{sec:m03}

\begin{figure}[t]
\begin{center}
%left, bottom, right, top
\resizebox{!}{5cm}{\includegraphics[trim=0 0 0 0 ]{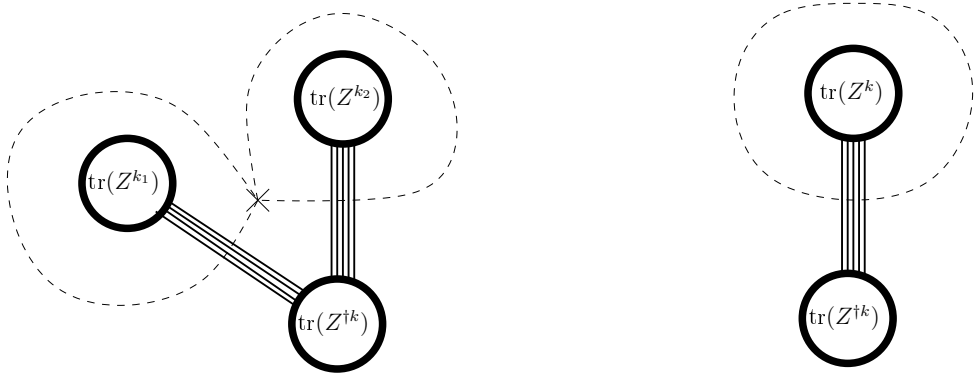}}
\caption[]{$Z$ model three-point function and two-point function on the sphere with bunched propagators.  The dual graph is drawn with a dashed line.}\label{fig:oomm-sph-3}
\end{center}
\end{figure}

The leading planar term of the three-point function in the $Z$ model is
\begin{equation}
  \corr{ \tr(Z^{\dagger k}) \tr(Z^{k_1})\tr(Z^{k_2})}_{\textrm{sphere}} = k_1k_2kN^{k-1} \label{eq:Zext3}
\end{equation}
where $k = k_1+k_2$.  This corresponds to the cut-and-join
operator $\S_{[2]}$ which splits $\tr(Z^{\dagger k})$ into two pieces,
cf. the analysis in \cite{1002.2099}, where there are two bunches of
homotopic propagators, see Figure \ref{fig:oomm-sph-3}.
\begin{figure}[p]
\begin{center}
%left, bottom, right, top
\resizebox{!}{3cm}{\includegraphics[trim=0 0 0 0 ]{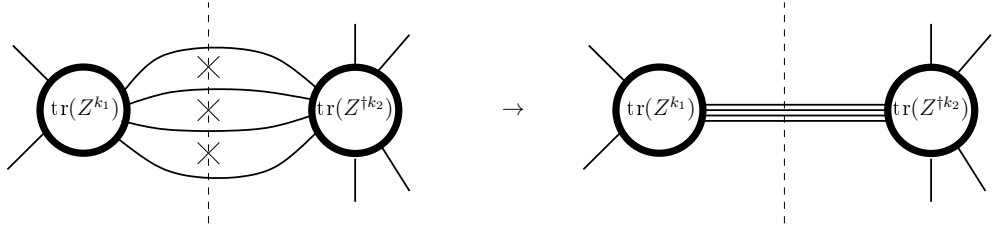}}
\caption[]{Replacement of 2-valent vertices of the $F$ model with plain $\tr(FF^\dagger)$ quadratic term (dashed lines), dual to faces bounded by parallel propagators in the $Z$ model (thin solid lines), by the propagator of the proper $F$ model \eqref{eq:FFprop}, which is dual to the parallel propagators of the $Z$  model bunched into an edge of the skeleton graph.  [All propagators are drawn in single-line notation here for ease of reading.]}\label{fig:oomm-parallel}
\end{center}
\end{figure}
The dual graph, drawn in Figure \ref{fig:oomm-sph-3} has one
four-valent vertex, two propagators and three faces (which correspond
to the old vertices).  Reading the term in the rescaled $F$ matrix model \eqref{eq:Fscaled}
\begin{align}
 \frac{N}{2}\corr{ \tr( FF^\dagger F
    F^\dagger)}&  =\frac{N}{2}\sum_{i_1,i_2,j_1,j_2}\corr{ F^{i_1}_{j_1} F^\dagger{}^{j_1}_{i_2}F^{i_2}_{j_2} F^\dagger{}^{j_2}_{i_1}} \nn \\
& =\frac{N}{2}\sum_{i_1,i_2,j_1,j_2}\left( \corr{ F^{i_1}_{j_1} F^\dagger{}^{j_1}_{i_2}}\corr{F^{i_2}_{j_2} F^\dagger{}^{j_2}_{i_1}}+ \corr{ F^{i_1}_{j_1}F^\dagger{}^{j_2}_{i_1} }\corr{F^{i_2}_{j_2} F^\dagger{}^{j_1}_{i_2}}\right)
\end{align}
Note that there are no non-planar terms in this correlator because of the configuration of the fields.\footnote{In a Gaussian Hermitian matrix model $\corr{\tr(M^4)}$ does however receive non-planar contributions.} Inserting the propagator \eqref{eq:Fscaledprop} and then Taylor expanding each one
\begin{align}
&\frac{N}{2} \sum_{i_1,i_2,j_1,j_2} \left(\frac{\delta^{i_1}_{i_2} \delta^{j_1}_{j_1}}{(a_{i_1} b_{j_1}-N)}\frac{\delta^{i_2}_{i_1} \delta^{j_2}_{j_2}}{(a_{i_2} b_{j_2}-N)}+ 
\frac{\delta^{i_1}_{i_1} \delta^{j_2}_{j_1}}{(a_{i_1} b_{j_2}-N)}
\frac{\delta^{i_2}_{i_2} \delta^{j_1}_{j_2}}{(a_{i_2} b_{j_1}-N)} \right) \nn \\
&=\frac{N}{2} \sum_{i,j_1,j_2}\sum_{k_1,k_2=1}^\infty \frac{N^{k_1+k_2-2}}{a_{i}^{k_1+k_2}b_{j_1}^{k_1}b_{j_2}^{k_2}} + \sum_{i_1,i_2,j}\sum_{k_1,k_2=1}^\infty \frac{N^{k_1+k_2-2}}{a_{i_1}^{k_1}a_{i_2}^{k_2}b_{j}^{k_1+k_2}} \nn \\
&= \frac{1}{2}\sum_{k_1,k_2=1}^\infty \left[ tr(A^{-k_1-k_2})  \tr(B^{-k_1})\tr(B^{-k_2})+\tr(A^{-k_1})\tr(A^{-k_2})tr(B^{-k_1-k_2}) \right]N^{k_1+k_2-1}\nn \\
& =   \sum_{k_1 <k_2} \left[ t_{k_1} t_{k_2} {\ov t}_{k_1+k_2}+t_{k_1+k_2} {\ov t}_{k_1}{\ov t}_{k_2}\right] k_1 k_2(k_1+k_2)N^{k_1+k_2-1} + \frac{1}{2!}\sum_{k_1} \left[ t_{k_1} t_{k_1} {\ov t}_{2k_1}+t_{2k_1} {\ov t}_{k_1}{\ov t}_{k_1}\right] 2k_1^3N^{2k_1-1} \nn
\end{align}
This agrees with the expectation from \eqref{eq:Zext3} where we get
contributions from the conjugate correlation function too.  Note that
the generating function splits into two pieces depending on
whether $k_1=k_2$, in which case we get a factorial from the
exponential in the $Z$ action \eqref{eq:Zmodel}.

\subsection{$\cM_{1,2}$}\label{sec:m12}

\begin{figure}[t]
\begin{center}
%left, bottom, right, top
\resizebox{!}{5cm}{\includegraphics[trim=0 0 0 0 ]{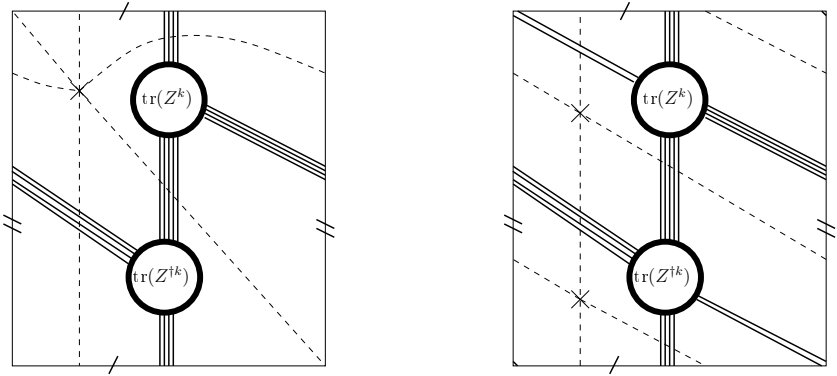}}
\caption[]{The two different bunchings of propagators with no crossing
  for the $Z$ model two-point function on the torus: three bunchings
  from $\S_{[3]}$ on the left and four bunchings from $\S_{[2,2]}$ on
  the right.  The dual graphs are drawn with dashed lines. The left
  dual graph can be considered as a degenerate case of the right graph
  when the two vertices of the dual graph on the right
  collide.}\label{fig:oomm-tor}
\end{center}
\end{figure}

The torus two-point function for the $Z$ model is
\begin{equation}
  \corr{ \tr(Z^{\dagger k}) \tr(Z^{k})}_{\textrm{torus}} = \bra{k}\left( \S_{[3]} + \S_{[2,2]}\right) \ket{k}N^{k-2} = k \left[ \binom{k}{3} + \binom{k}{4}\right]N^{k-2} \label{eq:2pttorus}
\end{equation}
Here we've used the cut-and-join notation of \cite{1002.2099}.  The different cut-and-join operators correspond to bunching  homotopic propagators into either 3 or 4 bunches, cf. Figure \ref{fig:oomm-tor} for the two possibilities.
The bunching of the propagators into three, the lefthand diagram in Figure \ref{fig:oomm-tor}, yields a dual graph with a single six-valent vertex, three edges and two faces.
 Reading the appropriate term in the $F$ matrix model we compute the non-planar torus term for the six-valent vertex
\begin{align}
 \frac{N}{3}\corr{\tr(F
    F^\dagger F
    F^\dagger F
    F^\dagger)}_{\textrm{torus}} & = \frac{N}{3}\sum_{i_1,i_2,i_3,j_1,j_2,j_3}  \corr{ F^{i_1}_{j_1} F^\dagger{}^{j_1}_{i_2}F^{i_2}_{j_2} F^\dagger{}^{j_2}_{i_3}F^{i_3}_{j_3} F^\dagger{}^{j_3}_{i_1}}_{\textrm{torus}} \nn \\
& =   \frac{N}{3}\sum_{i_1,i_2,i_3,j_1,j_2,j_3}   \corr{ F^{i_1}_{j_1}  F^\dagger{}^{j_2}_{i_3}} \corr{F^{i_3}_{j_3} F^\dagger{}^{j_1}_{i_2}}\corr{F^{i_2}_{j_2}  F^\dagger{}^{j_3}_{i_1}}\nn \\
& = \frac{N}{3}\sum_{i_1,i_2,i_3,j_1,j_2,j_3} \frac{\delta^{i_1}_{i_3} \delta^{j_2}_{j_1}}{(a_{i_1}b_{j_1}-N)}  \frac{\delta^{i_3}_{i_2} \delta^{j_1}_{j_3}}{(a_{i_3}b_{j_3}-N)}  \frac{\delta^{i_2}_{i_1} \delta^{j_3}_{j_2}}{(a_{i_2}b_{j_2}-N)} \nn \\
& =\frac{N}{3} \sum_{i,j} \frac{1}{(a_ib_j-N)^3}
\end{align}
Now Taylor expand, using the binomial theorem
\begin{align}
\frac{N}{3} \sum_{i,j} \sum_{k=3}^\infty \binom{k-1}{2}\frac{N^{k-3}}{a_{i}^kb_{j}^k}  
& = \frac{1}{3} \sum_{k=3}^\infty \binom{k-1}{2}\tr(A^{-k})\tr(B^{-k})N^{k-2} \nn \\
& =  \sum_{k=3}^\infty k\binom{k}{3} t_k {\ov t}_kN^{k-2}
\end{align}
This agrees with the expectation from \eqref{eq:2pttorus}.

The bunching of the propagators into four, the righthand diagram in Figure \ref{fig:oomm-tor}, yields a dual graph with two four-valent vertices, four edges and two faces\footnote{One might worry about disconnected non-planar graphs where each 4-vertex only contracts with itself.  Fortunately this diagram is not possible, cf. the $\cM_{0,3}$ example.}:
\begin{align}
& \frac{N^2}{2\cdot 2\cdot 2!} \corr{\tr(  F
    F^\dagger F
    F^\dagger)\;\;\tr(  F
    F^\dagger F
    F^\dagger)}_{\textrm{torus}} \nn \\
&\hspace{1.5cm} = \frac{N^2}{8}\sum_{i_1,i_2,i_3,i_4,j_1,j_2,j_3,j_4} \corr{ F^{i_1}_{j_1} F^\dagger{}^{j_1}_{i_2}F^{i_2}_{j_2} F^\dagger{}^{j_2}_{i_1}\;\;F^{i_3}_{j_3} F^\dagger{}^{j_3}_{i_4}F^{i_4}_{j_4} F^\dagger{}^{j_4}_{i_3}}_{\textrm{torus}} \nn
\end{align}
There are two relevant non-planar contractions
\begin{equation}
  \corr{F^{i_1}_{j_1} F^\dagger{}^{j_3}_{i_4}   } \corr{F^{i_2}_{j_2}F^\dagger{}^{j_4}_{i_3}  }\corr{F^{i_3}_{j_3} F^\dagger{}^{j_2}_{i_1} }\corr{ F^{i_4}_{j_4}  F^\dagger{}^{j_1}_{i_2}  }\;\; +\;\; \corr{ F^{i_1}_{j_1}F^\dagger{}^{j_4}_{i_3} }\corr{F^{i_2}_{j_2}  F^\dagger{}^{j_3}_{i_4}  }\corr{ F^{i_3}_{j_3}F^\dagger{}^{j_1}_{i_2}  } \corr{ F^{i_4}_{j_4}F^\dagger{}^{j_2}_{i_1} }
\end{equation}
These both give the same contribution so we get
\begin{align}
\frac{N^2}{4}\sum_{i,j} \frac{1}{(a_i b_j-N)^4} & = \frac{N^2}{4}\sum_{i,j} \sum_{k=4}^\infty \binom{k-1}{3}\frac{N^{k-4}}{a_{i}^kb_{j}^k}  \nn \\
& = \frac{1}{4} \sum_{k=4}^\infty \binom{k-1}{3}\tr(A^{-k})\tr(B^{-k})N^{k-2} \nn \\
& = \sum_{k=4}^\infty k\binom{k}{4} t_k {\ov t}_k N^{k-2}
\end{align}

\subsection{$\cM_{0,4}$ $1\to 3$}\label{sec:m04ex}

The leading planar term of the $1\to 3$ four-point function in the $Z$
model is
\begin{equation}
  \corr{ \tr(Z^{\dagger k}) \tr(Z^{k_1})\tr(Z^{k_2})\tr(Z^{k_2})}_{\textrm{sphere}} = k_1k_2k_3k(k-1)N^{k-2} \label{eq:Zext4}
\end{equation}
where $k = k_1+k_2+k_3$.  This result is made up from contributions
from the two cut-and-join operators $\S_{[2,2]}$ and $\S_{[3]}$,
corresponding to the lollipop diagram with 4 edges and the Y diagram
with 3 edges, as explained in \cite{0602226,1002.2099}.  The lollipop
comes from two 4-vertices of the $F$ model with a self-contraction
each.  The Y diagram comes from one of the planar ways of contracting
a single 6-vertex of the $F$ model.

\subsection{$\cM_{0,4}$ $2\to 2$}\label{sec:m04non}

For $k_1+k_2 = k_3+k_4$ and $\max\{k_i\} = k_1$ the planar connected $2\to 2$ four-point function is
\begin{equation}
  \corr{ \tr(Z^{\dagger k_1}) \tr(Z^{\dagger k_2})\;\; \tr(Z^{k_3})\tr(Z^{k_4})}_{\textrm{planar, connected}} = k_1k_2k_3k_4(k_1-1)N^{k_1+k_2-2} 
\end{equation}
For generic values of the $\{k_i\}$ the $Z$ graph has 4 bunched edges corresponding to $\S_{[2,2]}$
and the dual graph is the planar contraction of two 4-vertices of the
$F$ model with no self-contractions.  In a degenerate case where
$k_1-k_3 = k_2-k_4$ there are only 3 edges in the $Z$ model skeleton
graph corresponding to $\S_{[3]}$ and the dual graph has a single 6-valent vertex.

Note that this correlation function also receives
contributions at order $N^{k_1+k_2}$ and $N^{k_1+k_2-2}$ if $k_1 =
k_3$ from disconnected two-point functions.

\subsection{$\cM_{0,2}$}\label{sec:m02}

The $F$ model diagram dual to the planar
two-point function requires special treatment.  The dual graph (see
the righthand side of Figure \ref{fig:oomm-sph-3}) must be take using
the plain Gaussian propagator \eqref{eq:FFplain} with the quadratic
interaction terms in the faces between the parallel propagators of the $Z$ model.  The
$F$ model diagram corresponds to taking these quadratic interaction terms
daisy-chained with no self-contractions
\begin{align}
\frac{N^k}{k!} \corr{\left[ \tr(A^{-1}FB^{-1}F^\dagger)\right]^k}_{\textrm{plain, daisy}}  & = \frac{N^k}{k!} \corr{\prod_{p=1}^k \sum_{i_p,j_p} \frac{1}{a_{i_p}}\frac{1}{b_{j_p}} F^{i_p}_{j_p} F^{\dagger}{}^{j_p}_{i_p}}_{\textrm{plain,daisy}}  \nn \\
& = \frac{N^k}{k!} (k-1)!\prod_{p=1}^k \sum_{i_p,j_p} \frac{1}{a_{i_p}}\frac{1}{b_{j_p}}\corr{ F^{i_p}_{j_p} F^{\dagger}{}^{j_{p+1}}_{i_{p+1}}}_{\textrm{plain}} 
\end{align}
Here $i_{k+1} \equiv i_1$ and similarly for $j_{k+1}$.  There are
$(k-1)!$ completely equivalent ways of daisy-chaining the Wick
contractions.  Now inserting the plain propagator \eqref{eq:FFplain}
\begin{align}
 \frac{N^k}{k}\prod_{p=1}^k \sum_{i_p,j_p} \frac{1}{a_{i_p}}\frac{1}{b_{j_p}}\; \delta^{i_p}_{i_{p+1}} \delta^{j_{p+1}}_{j_p} & = \frac{N^k}{k} \sum_{i,j} \frac{1}{a_{i}^k}\frac{1}{b_{j}^k} = kN^k t_k {\ov t}_k =t_k {\ov t}_k \corr{\tr(Z^k) \;\; \tr(Z^{\dagger k})}_{\textrm{sphere}}
\end{align}
In this final step we have used the formula for the planar two-point function.

Note that because we are using the plain version of the $F$ model there is no sum over topologically identical $Z$ graphs with bunched propagators; here the $Z$ correlation functions must be calculated separately for each $k$.

\section{Examples for $C,D$ model}\label{sec:CDexamples}

In this section are calculations for the $C,D$ model
\eqref{eq:CD} with quartic vertex and propagator \eqref{eq:CDprop}.

\subsection{$\cM_{0,3}$}

For the specific 3-point function
\begin{equation}
  \corr{\tr(Z)\tr(Z)\tr(Z^{\dagger 2})}_{\textrm{sphere}} = 2N
\end{equation}
exactly one $C,D$ diagram contributes.  This diagram is the same as
(3) in Figure \ref{fig:oomm-graphd} except that there are only two
quartic vertices.  Proceeding with Einstein summation on
$e_k, f_k =1,\cdots N$ only
\begin{align}
&  \frac{1}{2!}\sum_{i_1,i_2;j_1,j_2}\corr{C_{e_1i_1}C^\dagger_{f_1i_1} D_{f_1j_1} D^\dagger_{e_1j_1}C_{e_2i_2}C^\dagger_{f_2i_2} D_{f_2j_2} D^\dagger_{e_2j_2}}\Big|_{Z,Z,Z^{\dagger 2}} \nn \\
& = \frac{1}{2} \sum_{i_1,i_2;j_1,j_2}\corr{C_{e_1i_1}C^\dagger_{f_1i_1}}\corr{D_{f_1j_1}D^\dagger_{e_2j_2}}\corr{ D^\dagger_{e_1j_1} D_{f_2j_2}   }\corr{ C_{e_2i_2}C^\dagger_{f_2i_2} } \nn \\
& = \frac{1}{2} \sum_{i_1,i_2;j_1,j_2}\frac{\delta_{e_1f_1}\delta_{i_1i_1}}{a_{i_1}}\frac{\delta_{f_1e_2}\delta_{j_1j_2}}{b_{j_1}}\frac{\delta_{e_1f_2}\delta_{j_1j_2}}{b_{j_1}}\frac{\delta_{e_2f_2}\delta_{i_2i_2}}{a_{i_2}} \nn \\
& = \frac{1}{2} \sum_{i_1,i_2;j} \frac{1}{a_{i_1}} \frac{1}{a_{i_2}} \frac{1}{b^2_{j}} N \;\; =\;\;\frac{1}{2!} t_1t_1 {\ov t}_2 \;2N
\end{align}

\subsection{$\cM_{1,2}$}

For the specific torus 2-point function
\begin{equation}
  \corr{\tr(Z^3)\tr(Z^{\dagger 3})}_{\textrm{torus}} = 3N \label{eq:CDtodo}
\end{equation}
the relevant torus $C,D$ diagram has three quartic vertices
\begin{align}
  \frac{1}{3!} \sum_{\{i_k\},\{j_k\}}\corr{C_{e_1i_1}C^\dagger_{f_1i_1} D_{f_1j_1} D^\dagger_{e_1j_1}C_{e_2i_2}C^\dagger_{f_2i_2} D_{f_2j_2} D^\dagger_{e_2j_2}C_{e_3i_3}C^\dagger_{f_3i_3} D_{f_3j_3} D^\dagger_{e_3j_3}}_{\textrm{torus}}
\end{align}
There are two ways of Wick contracting; one choice gives half of the result
\begin{align}
&  \frac{1}{3!} \sum_{\{i_k\},\{j_k\}} \corr{C_{e_1i_1}C^\dagger_{f_2i_2}}\corr{C^\dagger_{f_1i_1}C_{e_3i_3}}\corr{D_{f_1j_1}D^\dagger_{e_2j_2}} \corr{D^\dagger_{e_1j_1}D_{f_3j_3}}\corr{ C_{e_2i_2}C^\dagger_{f_3i_3}} \corr{D_{f_2j_2}D^\dagger_{e_3j_3}} \nn \\
& = \frac{1}{3!}\sum_{i,j} \frac{1}{a_{i}^3} \frac{1}{b_{j}^3} N \;\; = \;\;\frac{1}{2} t_3{\ov t}_3 \;3N
\end{align}

\section{Relation of $F$ model to $W_{\infty}$ model}\label{WinftoF}

In this section we show that the $W_{\infty}$ model
\cite{9505127,9208031} can be directly related to the $F$ model by a
change of variables.\footnote{The proof in this section was carried
  out in collaboration with Hanna Gr\"onqvist of the University of
  Helsinki.}  Take the $W_\infty$ model with $\n = -i\m = N$, so that
the $\log M$ term is tuned away, and expand the exponentiated
operators in the same way as in equation \eqref{eq:expexpansion}
\begin{align}
  \int [dM]^{H^+}_{N \times N}\; e^{-\tr(M) + \sum_{k=1}^\infty {\ov t}_k \tr[(MA^{-1})^k]} =  \int [dM]\;\; e^{-\tr(M)}  \sum_{l(R) \leq N}\chi_R(B^{-1}) \chi_R(MA^{-1}) 
\end{align}
The eigenvalues of $M$ must be positive semi-definite for
this integral to be well-defined.  Shortly we will see how this
condition is automatically implemented by the $F$ model.  Make the change of variables $M = UDU^\dagger$ where $U$ is unitary and $D$ is diagonal with eigenvalues $m_1, \dots m_N \geq 0$
\begin{align}
& \int [dU] \prod_{i=1}^N d m_i \;\Delta^2(m_i) \;e^{-\tr(D)}  \sum_{l(R) \leq N} \chi_R(B^{-1}) \chi_R(UDU^\dagger A^{-1}) \nn \\
&\hspace{1.5cm} = \int\prod_{i=1}^N d m_i\; \Delta^2(m_i) \;e^{-\tr(D)} \sum_{l(R) \leq N} \chi_R(B^{-1})  \frac{\chi_R(D)\chi_R( A^{-1})}{\dim_N R} \label{eq:toagree}
\end{align}
$\Delta(m_i)$ is the standard Vandermonde determinant.  In the final line we have used the integral \cite{Itzykson:1979fi}
\begin{equation}
  \int [dU]^U_{N\times N} \;\;\chi_R(UXU^\dagger Y) = \frac{\chi_R(X) \chi_R(Y)}{\dim_N R } \label{eq:IZ}
\end{equation}

We will now manipulate the $F$ model in a similar way to get the same
answer \eqref{eq:toagree}.  The complex matrix $F$ can be written with
two unitary matrices $U,W$ and a diagonal matrix $D$ \cite{Morris:1990cq}
\begin{equation}
  F = W\sqrt{D}U^\dagger \quad \quad F^\dagger  = U \sqrt{D} W^\dagger \label{eq:complexdecomp}
\end{equation}
The eigenvalues $m_1, \dots m_n \geq 0$ of the diagonal matrix $D$ are the real, non-negative eigenvalues of $FF^\dagger$.  The measure is then
\begin{align}
& \int [dF]^\C_{n\times n}\;\;e^{-\tr(FF^\dagger) +N \sum_{k=1}^\infty \frac{1}{k} \tr\left[\left( A^{-1}F
    B^{-1}F^\dagger \right)^{k} \right]} \nn \\
& =  \int [dU][dW]\prod_{i=1}^n d m_i \Delta^2(m_i) \;\;e^{-\tr( D) +N \sum_{k=1}^\infty \frac{1}{k} \tr\left[\left( A^{-1}W\sqrt{D}U^\dagger
    B^{-1} U \sqrt{D} W^\dagger\right)^{k} \right]}
\end{align}
Character expanding the exponential with \eqref{eq:expexpansion} and
using \eqref{eq:IZ} on the $[dW]$ integral
\begin{align}
&  \int [dU][dW]\prod_{i=1}^n d m_i \Delta^2(m_i)\;\; e^{-\tr( D)}
\sum_{l(R) \leq N} \chi_R(\bI_N)\; \chi_R( A^{-1}W\sqrt{D}U^\dagger
    B^{-1} U \sqrt{D} W^\dagger) \nn \\
& =   \int [dU]\prod_{i=1}^n d m_i \Delta^2(m_i)\;\; e^{-\tr( D)}
\sum_{l(R) \leq N}\dim_N R\; \frac{ \chi_R(A^{-1}) \chi_R(DU^\dagger
    B^{-1} U  )}{\dim_n R}
\end{align}
Next use \eqref{eq:IZ} on the $[dU]$ integral
\begin{align}
  \int \prod_{i=1}^n d m_i \Delta^2(m_i)\;\; e^{-\tr( D)}
\sum_{l(R) \leq N} \frac{\dim_N R\; \chi_R(A^{-1}) \chi_R(D)\chi_R(B^{-1})}{\dim_n R \dim_n R} \label{eq:Fsd}
\end{align}
This already agrees with \eqref{eq:toagree} if we choose $n=N$.  For
the case $n\geq N$, compare \eqref{eq:Fsd} with the character expansion of the $F$
model \eqref{eq:expansion2} to see that
\begin{align}
  \int \prod_{i=1}^n d m_i \Delta^2(m_i)\;\; e^{-\tr( D)}\;
 \chi_R(D) = \left[\dim_n R\right]^2
\end{align}
Inserting this into \eqref{eq:toagree} we get agreement with
\eqref{eq:expansion2}.

\section{Complex matrix model duality for two (or more)}\label{sec:2more}

In this section we sketch how the duality might work for a $V$-type
model with two $N\times N$ complex matrices $X,Y$, corresponding to
two of the three complex scalars of free 4d $\cN=4$ SYM:
\begin{align}
&  \int [dX]^\C[dY]^\C\; e^{ -\tr(XX^\dagger) - \tr(YY^\dagger) +\sum_{\m_1,\m_2,[\a]} t_{\{ \m_1,\m_2,[\a]\}} \tr(\a\,X^{ \m_1}Y^{ \m_2} ) + \sum_{\m_1,\m_2,[\a]} {\ov t}_{\{ \m_1,\m_2,[\a]\}}\tr(\a\,X^{ \dagger \m_1}Y^{ \dagger \m_2} ) } \nn 
\end{align}
The sum is over all the holomorphic single-trace operators built
out of $\m_1$ $X$'s  and $\m_2$ $Y$'s.  $\a$ is a single $k$-cycle $\a \in [k] \subset
S_{\m_1+\m_2}$ where $k = \m_1+\m_2$.  The trace with a permutation is defined by
\begin{equation}
  \tr(\a\,X^{ \m_1}Y^{ \m_2} ) = X^{i_1}_{i_{\a(1)}} \cdots X^{i_{\m_1}}_{i_{\a(\m_1)}} Y^{i_{\m_1+1}}_{i_{\a(\m_1+1)}} \cdots Y^{i_{\m_1+\m_2}}_{i_{\a(\m_1+\m_2)}}
\end{equation}
It is unique up to conjugation $\a \sim \rho^{-1} \a \rho$ for $\rho
\in S_{\m_1}\times S_{\m_2}$ so we only sum over conjagacy classes
$[\a]$ for this relation.  The couplings $t, {\ov t}$ can be encoded
in a generalised Kontsevich-Miwa transformation
\begin{align}
   t_{\{ \m_1,\m_2,[\a]\}} & = \tfrac{1}{|\Sym(\a)\cap S_{\m_1}\times S_{\m_2}|}\tr(\a\; A^{\m_1} C^{\m_2}) \nn \\
   {\ov t}_{\{ \m_1,\m_2,[\a]\}} & = \tfrac{1}{|\Sym(\a)\cap S_{\m_1}\times S_{\m_2}|}\tr(\a\; B^{\m_1} D^{\m_2}) 
\end{align}
The matrices $A,B,C,D$ do not commute and are not diagonalisable,
unlike the single complex matrix case. For a single
cycle $\Sym(\a) \cong \Z_k$. Some examples:
\begin{align}
 \begin{array}{rlcrl} t_{\tr(X^k)} & = \frac{1}{k} \tr(A^k) &\quad& t_{\tr(X^2Y^2)} & = \tr(A^2C^2) \nn \\
 t_{\tr(Y^k)} & = \frac{1}{k} \tr(C^k) & \quad& t_{\tr(XYXY)} & = \frac{1}{2}\tr(ACAC)
 \end{array}
\end{align}
To get the dual model of $F$-type the techniques of Section
\ref{sec:proof} using integration in-out look inapplicable.  Character
expansions may work.  A guess based on graph duality is
\begin{align}
  \int [dF]^\C[dG]^\C \;e^{ -\tr(FF^\dagger)-\tr(GG^\dagger)+ \sum_{k_i,[\a]}s_{\{k_i,[\a] \}}  \tr(\a\; (AFBF^\dagger)^{k_1} (AFDG^\dagger)^{k_2}(CGBF^\dagger)^{k_3}(CGDG^\dagger)^{k_4}) }
\end{align}
Each $F$ propagator is transverse to an $X$ propagator and similarly
for $G$ and $Y$.  This guess has been checked for very simple
two- and three-point functions.  Here $\a$ is a single cycle permutation in
$S_{k}$ where $k = \sum_{i=1}^4 k_i$ and the coupling is defined
\begin{align}
  s_{\{k_i,[\a] \}} = \tfrac{N}{|\Sym(\a)\cap \prod_iS_{k_i}|}
\end{align}

\end{appendix}

\end{document}